\documentclass[10pt, twocolumn]{IEEEtran}
\usepackage{amsmath}
\usepackage{amssymb}
\usepackage{cite}
\usepackage{color}
\usepackage{float}
\usepackage{mathtools}
\usepackage{microtype}
\usepackage{setspace}
\usepackage[caption=false]{subfig}
\usepackage[para]{footmisc}
\restylefloat{table}

\begin{document}
\title{A Deep Reinforcement Learning Framework for Contention-Based Spectrum Sharing}
\author{Akash Doshi, Srinivas Yerramalli, Lorenzo Ferrari, Taesang Yoo, Jeffrey G. Andrews
\thanks{Akash Doshi and Jeffrey G. Andrews are with The University of Texas at Austin, TX 78712 (e-mail: akashsdoshi@utexas.edu, jandrews@ece.utexas.edu). Srinivas Yerramalli, Lorenzo Ferrari and  Taesang Yoo are with  Qualcomm Research Center, San Diego, CA 92121 (e-mail: syerrama, lferrari, taesangy@qti.qualcomm.com). This work commenced while the first author was an intern at Qualcomm.}}

\maketitle 
\normalsize
\begin{abstract}
The increasing number of wireless devices operating in unlicensed spectrum motivates the development of intelligent adaptive approaches to spectrum access. We consider decentralized contention-based medium access for base stations (BSs) operating on unlicensed shared spectrum, where each BS autonomously decides whether or not to transmit on a given resource.  The contention decision attempts to maximize not its own downlink throughput, but rather a network-wide objective.  We formulate this problem as a decentralized partially observable Markov decision process with a novel reward structure that provides long term proportional fairness in terms of throughput. We then introduce a two-stage Markov decision process in each time slot that uses information from spectrum sensing and reception quality to make a medium access decision. Finally, we incorporate these features into a distributed reinforcement learning framework for contention-based spectrum access. Our formulation provides decentralized inference, online adaptability and also caters to partial observability of the environment through recurrent Q-learning.  Empirically, we find its maximization of the proportional fairness metric to be competitive with a genie-aided adaptive energy detection threshold, while being robust to channel fading and small contention windows. 
\end{abstract}


\section{Introduction}
Spectrum sharing attempts to allow different transmitters to operate on the same allocated resource (spectrum/time) in a fair manner, while also providing high throughput. Because each transmitter cannot be cognizant of the complete system state, an optimal decentralized policy for all transmitters is infeasible \cite{zhao2007decentralized}.  Centralized approaches require excessive real-time overhead messaging between the transmitters, and obtaining the solution is known to be NP-hard, since all the transmissions and throughputs are coupled, and has been shown in \cite{zheng2005collaboration} to be equivalent to graph coloring. We assume that transmitters will operate in a decentralized fashion, such that there is no direct exchange of messages between them before making a decision in a given time slot. Instead, each transmitter will perform spectrum sensing and also utilize local side information before it accesses the channel.

\subsection{Motivation}
To alleviate spectrum constraints, the usage of Long Term Evolution (LTE) in the unlicensed 5 GHz band (LTE-U) was introduced in 2014 by Qualcomm \cite{qualcommlteu2014}. Subsequently, 3GPP \cite{3gpp.36.889} standardized License Assisted Access (LAA) for LTE - with LTE-LAA, a licensed carrier, used for control signalling, can be thought of as an anchor that stays connected as unlicensed carriers, used to transport data, are added to or dropped from the combination of carriers in use between a device and the network. The approach adopted to access the unlicensed spectrum in LAA is known as Listen-Before-Talk (LBT) \cite{3gpp.36.889}\cite{etsi} and required each transmitter to perform a Clear Channel Assessment (CCA) before accessing spectrum. In other words, a base station (BS) is allowed to transmit on a channel only if the energy level in the channel is less than the CCA threshold level for the duration of the CCA observation time \cite{etsi}. The duration of the CCA observation time can be fixed or randomized depending on the mode of LBT being employed \cite{3gpp.38.889}. The 5G New Radio Unlicensed (NR-U) is a successor to LTE-LAA that is designed to operate in the 5 and 6 GHz bands and encompasses both license-assisted and standalone access i.e. without any anchor in licensed spectrum. The 3GPP study on NR-U \cite{3gpp.38.889} mentions that LTE-LAA LBT would be the starting point of the design for the 6GHz band. Hence, in this paper, we utilize spectrum sensing via LBT as a baseline to benchmark the performance of the algorithm we develop to improve UE throughput under NR-U in the 6 GHz band.

\begin{figure}
    \centering
    \includegraphics[width=1.9in,height=1.1in]{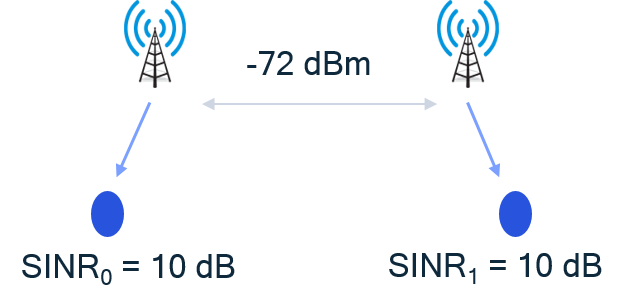}
    \hspace{0.3in}
    \includegraphics[width=1.2in,height=1.1in]{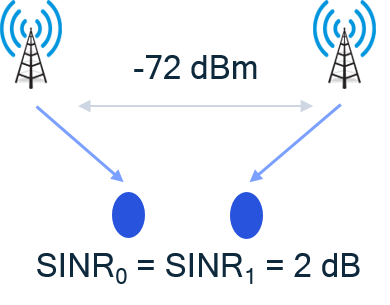}
    \caption{Two simple layouts of 2 BS's and 2 UE's, depicting the need for an adaptive ED threshold at the BS.}
    \label{fig:layouts}
\end{figure}

The usage of a given CCA threshold level, also referred to as an energy detect (ED) threshold in NR-U, as a channel access policy can often negatively impact the throughput. For instance, consider the toy scenario depicted in Fig. \ref{fig:layouts} involving only BS to User Equipment (UE) downlink (DL) transmissions as indicated. In the scenario on the left, there is minimal interference across links, hence simultaneous DL transmission is possible, while in the other, there is strong interference across links, so the DL transmissions should be time multiplexed. A common ED threshold at the BSs cannot account for both scenarios, since "energy detect" at the BSs does not actually reflect the quality of reception (SINR) at the UE. Moreover, an ED threshold-based medium access mechanism is incapable of adapting to varying path gains brought about by fading and mobile UE positions. 

Hence what we need is a medium access algorithm that is able to jointly process (i) information about reception quality at the UE from the \emph{previous} time slot(s), which can be acquired using standard channel quality indicator (CQI) feedback as well as monitoring the throughput achieved to that UE, and (ii) energy levels \emph{currently} sensed at the BS, which reflect the current interference state but not at the UE itself.   Thus, we can track past interference statistics at the desired location (the receiver), and observe the current interference at the transmitter. Based on these observations, a BS would have to decide whether or not to transmit in the current time slot. This must be done without direct awareness of the actions of the other BSs, or knowledge of the long term throughput patterns at UEs that the BS does not serve.  

Multi-agent reinforcement learning (RL) techniques \cite{bucsoniu2010multi} provide a principled approach to incorporating environment dynamics and designing agents that are trained to take decisions in the face of uncertainty. This motivates us to design a novel distributed reinforcement learning-based approach for medium access at the BS that uses information about the past quality of signal reception at the UE it serves in combination with channel sensing to adapt its medium access policy.  

\subsection{Related Work and Approaches}
Contention-based medium access is a fundamental and classic problem in wireless systems. The notion of LBT and deciding on whether the channel is clear based on an energy threshold was introduced in the earliest IEEE 802.11 WiFi protocols, and still is used \cite{patriciello2020nr}, with very little subsequent improvement. In systems where data is transmitted in time slots of fixed length, referred to as Frame Based Equipment (FBE) by \cite{etsi}, each time slot is divided into a contention period and a data transmission phase (refer Fig. \ref{fig:prob_statement}). During the contention period (also known as I-CCA in \cite{3gpp.38.889}), the BS senses the channel and if it remains idle (based on the ED threshold), the BS transmits data for a fixed time period (a maximum of $9$ ms) and then contends for the channel again. Later variants of LBT, such as Cat3 and Cat4 \cite{3gpp.38.889}, introduced an extended CCA period (E-CCA) in which a BS senses the channel for a further random duration if the channel remained idle during I-CCA. To determine this random duration, a BS will draw a random counter $\theta$ between $0$ and $\mathrm{CWS} - 1$ (where CWS denotes the contention window size e.g. CWS = 6 in Fig. \ref{fig:prob_statement}). It will then decrement $\theta$ by 1 every $5~\mu \mathrm{s}$, and transmit after $\theta$ hits $0$ only if the channel remained idle for the entire duration. In this paper, we adopt a contention-based channel access scheme that is a simplified version of Cat3 LBT in NR-U\cite{3gpp.38.889}.

IEEE 802.11 also supports an optional collision reduction scheme known as RTS/CTS, such that one node (e.g. a BS) sends a Request-to-Send (RTS) frame if it senses a clear channel, and prior to sending data. In response, the receiving node (e.g. a UE) sends a Clear-to-Send (CTS) frame if the received Signal-to-Interference Ratio (SINR) of the RTS frame was above a certain threshold \cite{jamil2015efficient}. However, RTS-CTS can inhibit potentially successful transmissions, and introduces significant additional overhead and latency \cite{sobrinho2005rts}.  Several empirical Media Access Control (MAC) layer optimizations have been proposed subsequently, including dynamic blocking notification schemes \cite{zhang2006ebn} \cite{chong2014dynamic}, but they all suffer from similar drawbacks.

A more recent methodology is to apply multi-agent reinforcement learning to design state-based policies that can improve the performance of unlicensed spectrum sharing. Initial work such as \cite{li2010multiagent} and \cite{lunden2011reinforcement} utilized function approximation and tabular listing of $\mathcal{Q}$-values for spectrum access in cognitive radio systems, which was not scalable to large state spaces. More recent work such as \cite{tonnemacher2018opportunistic} and \cite{tonnemacher2019machine} began to utilize deep $\mathcal{Q}$-learning \cite{mnih2015human} to either chose an action that adapts the ED threshold to the BS queue length or chooses the optimal subcarrier. Most recently, \cite{liang2019spectrum} and \cite{li2019multi} used deep $\mathcal{Q}$-learning to improve sum rate in vehicle-to-vehicle (V2V) and device-to-device (D2D) communication, while \cite{chang2018distributive} additionally employed reservoir computing, a special type of recurrent neural network to combat partial observability. However, these papers are all focused on supporting underlay communications and not on optimizing medium access of the primary spectrum user. A key feature of the 6 GHz band is that no unlicensed devices currently operate there and hence NR-U channel access protocols designed for it would not be constrained by primary spectrum users \cite{naik2020next}.

In \cite{naparstek2018deep}, game theoretic principles are used in designing a RL reward function that maximizes the number of successful transmissions in a distributed setting. However, they require receipt of an ACK as part of their algorithm, consider all links at the same SNR and do not utilize information from spectrum sensing. Most recently, \cite{naderializadeh2021resource} presented a robust and scalable distributed RL design for radio resource management to mitigate interference. None of these papers thus far have attempted to model the asynchronous nature of the decisions made by the transmitters owing to contention. Moreover \cite{naderializadeh2021resource} assumes that APs exchange messages in every time slot on backhaul links to provide remote observations required for their policy, which may not be possible in a practical deployment, particularly across different operators or systems.

\subsection{Contributions}
We design a novel distributed reinforcement learning approach to optimizing medium access that is aimed at improving the current LBT-based approach in NR-U \cite{3gpp.38.889}, by constraining our problem to a contention-based access mechanism. We employ the paradigm of \textit{centralized learning} with \textit{decentralized execution}, such that each BS will decide whether and how to transmit based only on its own observations of the system. Our technical contributions are now summarized.

\textbf{Formulating Medium Access as a \textit{DEC-POMDP}}.  In a practical deployment, a BS will only have access to delayed copies of the parameters of the UEs it serves in each time slot and will not be able to directly observe the action of all neighbouring BSs. Moreover, there is no central controller that can determine the action of each BS. We formulate medium access decisions as a \textit{decentralized partially observable Markov decision process (DEC-POMDP)} \cite{bernstein2002complexity} by incorporating these key features of a practical deployment.

\textbf{Adapting a Medium Access \textit{DEC-POMDP} to Contention}. In each time slot, a BS can either be transmitting to a UE or waiting in the contention queue. This motivates a 2-state transition diagram, such that a BS can either be in the data transmission state (until the end of the time slot) or the contention state. We define each of the states in accordance with the information available at a BS in each time slot. We formulate a reward structure associated with each state transition, such that maximization of the sum of the rewards accumulated over time provides long term proportional fairness (PF) of the throughput delivered to all UEs.

\textbf{Solving a Medium Access \textit{DEC-POMDP} using an Independent DQN}. We adapt a distributed reinforcement learning strategy that combines Deep $\mathcal{Q}$-networks (DQN) with \textit{independent Q-learning} \cite{tampuu2017multiagent} and adds recurrency to model partial observability \cite{hausknecht2015deep}. Inspired by the inter-agent communication framework of \cite{foerster2016learning} that provides for end-to-end learning across agents in a decentralized setting, we exploit the inter-BS energies detected along with the local action-observation history to successfully train two DQNs at each BS using a centralized training procedure that provides for decentralized inference.  We show that this approach, after a modest number of iterations, achieves maximization of the PF metric competitive with a \textit{genie-aided adaptive ED threshold} that unrealistically presumes knowledge of the UE locations to chose the optimal energy threshold.

\vspace{2mm}
The paper is organized as follows. In Section \ref{sec:PF_scheduler}, the system model is given, and a mathematical formulation of the problem statement is provided in terms of proportional fairness, along with a solution that requires a central controller.  A realistic decentralized inference framework for medium access is developed in Section \ref{sec:dec_ma}, which is adapted to a distributed Reinforcement Learning (RL) framework in Section \ref{sec:ind_dqn}. The simulations and detailed results are presented in Section \ref{sec:sim_details} and \ref{sec:Results} respectively, followed by the conclusions and possible future directions in Section \ref{sec:conc}.

\section{Problem Statement and System Model} \label{sec:PF_scheduler}

\subsection{An Overview}

We consider a downlink cellular deployment of $N$ BSs, each BS having at least one active UE desiring data transmission. Assume, as in LAA, that the DL transmissions for all BSs occur on the same shared sub-band of unlicensed spectrum, while the uplink (UL) transmission of CQI and other control information takes place on separate licensed channels for each BS-UE link. We separate the scheduler block that determines the served UE per BS from the medium access block, and focus only on medium access control (MAC). The BS transmits at the maximum MCS (Modulation and Coding Scheme) allowable at that SINR, assuming that the UE throughput can be approximated by the Shannon capacity. Assuming that the same UE is scheduled for reception for $L$ consecutive time slots and each BS transmits at a constant power, the MAC algorithm at each BS has to decide whether or not to transmit to the UE in each time slot. Moreover, we assume that the BSs are backlogged with sufficient packets such that the BS always has traffic to be delivered to its scheduled UE, and hence always participates in contention. This is the worst-case scenario: if a BS does not choose to participate in contention, it improves throughput for other cell's UEs.
\begin{figure}
    \centering
    \includegraphics[width=3.1in]{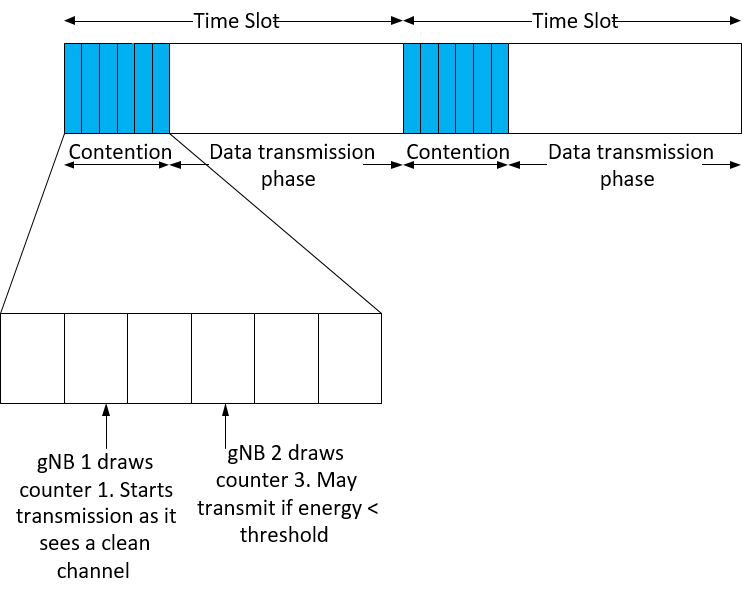}
    \caption{Contention-based access for FBE to unlicensed spectrum in BS-UE DL transmissions}
    \label{fig:prob_statement}
\end{figure}

We consider a simplified contention-based access mechanism for FBE as shown in Fig. \ref{fig:prob_statement}. Each BS draws a random counter in $\{0, \ldots, \mathrm{CWS}-1\}$ at the start of a time slot. For simplicity, we assume that each BS draws a unique counter such that there is no collision among counters drawn at different BSs in the contention process\footnote{This assumption is for the simplicity of the state-action transition; the algorithm itself does not require such uniqueness (refer to Section \ref{subsec:nuc})}. Hence, the contention window (CW) is of length at least equal to the number of BSs $N$ in the layout, i.e. $\mathrm{CWS} \geq N$. When this counter expires, the BS ascertains if the channel is clear before transmitting as shown in Fig. \ref{fig:prob_statement}. If the channel is clear, the BS transmits a unique preamble (which can be used by other BSs to identify the transmitting BS) for the remainder of the contention phase, followed by data symbols from the beginning of the data transmission period. The objective of each BS is to maximize the long-term throughput seen by the UE. We now formulate this mathematically.

\subsection{Mathematical Formulation}
A single UE is scheduled per contention slot per BS.  For each BS $i$, $a_i \in \{0,1\}$ denotes the action chosen by the BS to transmit or not, with transmission denoted by $1$ and the action vector of all BSs given by $\mathbf{a}$.  Considering single-input single-output (SISO) communication between BSs and UEs in a single sub-band, we denote the channel between the BS $i$ and UE $j$ by $h_{ij}$, and the channel between BS $i$ and BS $j$ by $h_{ij}'$ where $i$ and $j$ are drawn from $\{1,\ldots N\}$. The respective path gains are $g_{ij} = |h_{ij}|^2$ and $g_{ij}' = |h_{ij}'|^2$. For each UE $j$, $S_j$ denotes the desired signal power, $I_j$ the total interference power, $R_j$ the data rate experienced and $\overline{X}_j$ the exponentially smoothed average rate seen by the UE, with the vector containing these terms for all UEs being denoted by $\mathbf{S}$, $\mathbf{I}$, $\mathbf{R}$ and $\mathbf{\overline{X}}$ respectively. Denote the noise variance for DL receptions at the UE by $\sigma_{\mathrm{UE}}^2$, and at the BS by $\sigma_{\mathrm{BS}}^2$. 

Consider transmissions on a DL slot from BSs to UEs. Assuming each BS $i$ transmits at a fixed power $P_t$, the received signal at each UE is given by
\begin{equation}
    y_j = \sqrt{P_t}h_{jj}a_j + \sum_{i \neq j}\sqrt{P_t}h_{ij}a_i + z_j,
\end{equation}
where $z_j \sim \mathcal{CN}(0,\sigma_{\mathrm{UE}}^2)$, with $\mathcal{CN}$ denoting the complex normal distribution and $\sigma_{\mathrm{UE}}^2$ the noise variance at the UE. Then the SINR  measured at UE $j$ at the end of data transmission in time slot $n$ is given by
\begin{equation} \label{eq:SINR}
    \mathrm{SINR}_j[n] = \frac{g_{jj}[n]P_ta_j[n]}{\sigma_{\mathrm{UE}}^2 + \sum_{i=1,i \neq j}^{i=N}g_{ij}[n]P_ta_i[n]} = \frac{S_j}{\sigma_{\mathrm{UE}}^2 + I_j}.
\end{equation}
It should be noted that while we restrict ourselves in this exposition to SISO on a single sub-band, the definition of $g_{ij}$ could easily be generalized to simply represent the received signal power at UE $j$, which could then be utilized to incorporate both MIMO and frequency-selective channels. We assume the throughput $R_j[n]$ experienced by the UE $j$ in time slot $n$ to be ideal and for a bandwidth W is approximated by the Shannon capacity formula
\begin{equation} \label{eq:Rate}
    R_j[n] = W\log_2({1 + \mathrm{SINR}_j[n]}).
\end{equation}
Defining the utility function as
\begin{equation} \label{eq:utility_func}
    U(\mathbf{\overline{X}}[n]) = \sum_{j=1}^{N}\log(\overline{X}_j[n]),
\end{equation}
our objective is to have the BSs choose an action vector $\mathbf{a}[n]$ in each time slot $n$ that provides for long term proportional fairness. This means that we want to maximize the long term average rate $\lim_{n\rightarrow \infty} \overline{X}_j[n]$ of each UE $j$, where $\overline{X}_j[n]$ is the exponentially smoothed average rate seen by UE $j$ up to time $n$ and is given by
\begin{equation} \label{eq:avg_rate}
    \overline{X}_j[n] = (1-1/B)\overline{X}_j[n-1] + (1/B)R_j[n],
\end{equation}
with $B > 1$ being a parameter which balances the weights of past and current transmission rates. Mathematically, in \cite{kelly1998rate}, this is proved to be equivalent to maximizing the utility function $U(\mathbf{\overline{X}}[n])$ as defined in \eqref{eq:utility_func} for $n \rightarrow \infty$. In \cite{wang2010scheduling}, it is proven that maximizing $U(\mathbf{\overline{X}}[n])$ can be achieved iteratively through slot-by-slot BS scheduling. At each time slot, the iterative BS scheduler has to find a rate vector $\mathbf{R}^*[n]$ such that
\begin{equation} \label{eq:iterative_scheduler}
    \mathbf{R}^*[n] = \underset{\mathbf{R}[n]}{\mathrm{arg\ max\ }} \sum_{j=1}^{N} \frac{R_j[n]}{\overline{X}_j[n]},
\end{equation}
where $\mathbf{R}[n] = \langle R_1[n],\ldots R_N[n]\rangle$. For a given time slot $n$, this is equivalent to finding the action vector $\mathbf{a}[n]$ that maximizes the summation in \eqref{eq:iterative_scheduler}. This is a combinatorial search over $2^N$ possible binary action vectors. However, it is important to note that this solution requires a central controller and hence is not realizable in any practical decentralized deployment. We relax this assumption next.

\section{A Decentralized Medium Access Formulation} \label{sec:dec_ma}
As described in Section \ref{sec:PF_scheduler}, the PF-based BS scheduler requires awareness of all the link strengths and average rate of all UEs, such that it can instruct all BSs on the decision to make in each time slot. In reality, we have distributed control by $N$ BSs, with the decision of each BS determined by its own observation of the environment. As a consequence, each BS is usually not aware of the decisions of all other BSs in the given time slot. Moreover, each BS $i$ is cognizant of only the average rate $\overline{X}_i$ of the UE it serves, and not that of all the UEs, $\mathbf{\overline{X}}$. Finally, in a fading environment, each BS $i$ will not be aware of the entire $N\times N$ interference channel matrix $\mathbf{G} = \{g_{ij}\}$, however it can utilize the signal and interference power $S_i$ and $I_i$ experienced by the UE it serves in the previous data transmission, which is fed back to the BS in uplink, as indicators of the magnitude of the channel coefficients. This motivates us to formally define the MAC for BSs operating on shared spectrum under the framework of a \textit{DEC-POMDP} \cite{bernstein2002complexity}.

\subsection{Defining a DEC-POMDP} \label{subsec:dec_pomdp}
The medium access \textit{DEC-POMDP} can be defined as the 8-tuple $\langle \mathcal{B}, \mathcal{S},\mathcal{A},T,\mathcal{R},\Omega,\mathcal{O},\gamma \rangle$ where $\mathcal{B}$ is a set of $N$ BSs (agents), $\mathcal{S}$ is the state space, $\mathcal{A}=\times_i A_i$ ($\times_i$ denotes a Cartesian product) is the joint action space, and $\Omega =\times_i \omega_i$ is the joint observation space of all BSs. At each time step $n$, each BS executes an action $a_i \in A_i = \{0,1\}$, causing the environment state $\mathbf{s} \in \mathcal{S}$ to transition to $\mathbf{s}'$ with probability  $P(\mathbf{s}'|\mathbf{s},\mathbf{a}) = T(\mathbf{s},\mathbf{a},\mathbf{s}')$. Each BS then receives observation $o_i \in \omega_i$ based on the joint observation function $\mathcal{O}(\mathbf{o}| \mathbf{s}',\mathbf{a})$ where $\mathbf{o} = \langle o_1, \ldots, o_N \rangle$. Define the local action-observation history at timestep $n$ as $\vec{o}_i[n] = (o_i[1], \ldots, o_i[n])$, where $\vec{o}_i[n] \in \vec{\omega}_i[n]$. Single-agent policies $\pi_i: \vec{\omega}_i[n] \rightarrow A_i$ select an action for each BS $i$, with the joint policy being denoted by $\mathbf{\pi} = \langle \pi_1, \ldots, \pi_N \rangle$. All BSs receive a single common reward $r = \mathcal{R}(\mathbf{s'},\mathbf{a})$ at the end of each time slot. The exact formulations for $\vec{o}_i[n]$ and $r$ will be described in Section \ref{subsec:med_acc_dec_pomdp}, however it should be noted that we assume centralized computation of $r$ offline, since online computation of a centralized reward would require communication among BSs on backhaul links. Finally, the objective is to learn the optimal joint policy $\mathbf{\pi}^* = \langle \pi_1^*, \ldots, \pi_N^* \rangle$ that maximizes the expected cumulative reward $\sum_{n=0}^{L} \gamma^n r[n]$ over a finite time horizon of $L$ consecutive time slots for some discount parameter $\gamma \in (0,1)$. We will now define the state space $\mathcal{S}$, transition function $T(\mathbf{s},\mathbf{a},\mathbf{s}')$, reward function $\mathcal{R}(\mathbf{s},\mathbf{a})$ and joint observation space $\Omega$ used for modelling medium access. The notation employed has been summarized in Table \ref{tab:decpomdp_notation}.
\begin{table}
    \caption[\textit{DEC-POMDP} Notation] {\textit{DEC-POMDP} Notation}
	\label{tab:decpomdp_notation}
	\centering
	\begin{tabular}{ |p{0.5cm}|p{6.4cm}|}
		\hline
		$a_i$ & Action chosen by BS $i$\\\hline
		$A_i$ & Action space of BS $i$ $= \{0,1\}$\\\hline
		$o_i$ & Local observation received by BS $i$\\\hline
		$\omega_i$ & Observation space of BS $i$
		\\\hline
		$\vec{o}_i$ & Local action-observation history of BS $i$
		\\\hline
		$\vec{\omega}_i$ & Action-observation history space of BS $i$
		\\\hline
		$\pi_i$ & Policy used to choose action $a_i$ at BS $i$ \\\hline
		$r$ & Single common reward distributed to all BSs \\\hline
	\end{tabular}
\end{table}

\subsection{Formulating Medium Access as a DEC-POMDP} \label{subsec:med_acc_dec_pomdp}
Denote by $\mathbf{G}[n] = \{g_{ij}[n]\}$ the downlink $N \times N$\footnote{In reality, the BS cluster size $N$ would be determined by the number of BS transmissions resulting in non-negligible interference at a given site. However, in this work, we consider a fixed $N$ throughout.} interference channel matrix in time slot $n$. To solve \eqref{eq:iterative_scheduler}, in each time slot, we would require knowledge of the action of all BSs $\mathbf{a}[n]$, the average rates at the beginning of the time slot $\mathbf{\overline{X}}[n-1]$ and the interference matrix $\mathbf{G}[n]$. Hence, the state of the system $\mathbf{s}[n] \triangleq \langle \mathbf{\overline{X}}[n-1],\mathbf{G}[n]\rangle$. The transition function $ T(\mathbf{s}[n],\mathbf{a},\mathbf{s}[n+1])$ is given by combining \eqref{eq:SINR}, \eqref{eq:Rate} and \eqref{eq:avg_rate}. Hence, with $\mathbf{\overline{X}} = \langle \overline{X}_1, \ldots \overline{X}_N \rangle$, we have
\begin{align} 
    &\overline{X}_j[n] = \Big(1-\frac{1}{B}\Big)\overline{X}_j[n-1] + \frac{1}{B} R_j[n], \label{eq:rate_update}\\
    &R_j[n] = \log_2\Bigg(1+\frac{g_{jj}[n]a_j[n]}{\sigma^2 + \sum_{i \neq j}g_{ij}[n]a_i[n]}\Bigg)\\
    &g_{ij}[n+1] = f(g_{ij}[n]) + n_{ij}, \label{eq:path_gain_update}
\end{align}
where $\sigma^2 = \sigma_{\mathrm{UE}}^2/P_t$ and $n_{ij}$ is the noise term. We drop the bandwidth scaling factor $W$ since it would just add a constant term to $U(\mathbf{\overline{X}}[L])$. Note how $\overline{X}_j[n]$ also depends on the action $a_i[n]$ of all BSs $i \neq j$ in \eqref{eq:rate_update}. The probabilistic nature of the transition from $\mathbf{s}[n]$ to $\mathbf{s}[n+1]$ is captured by the noise term $n_{ij}$ in \eqref{eq:path_gain_update}. The function $f$ in \eqref{eq:path_gain_update} is used to convey a first order Markov chain representation of a channel model, known to be sufficient for modelling very slow fading channels \cite{tan1998first}. Given $L$ time slots, while long term PF-based BS scheduling would seek to maximize $U(\mathbf{\overline{X}}[L])$, solving the medium access \textit{DEC-POMDP} entails maximization of $\sum_{n=0}^{L} \gamma^n r[n]$. Hence we would like to have 
\begin{equation} \label{eq:rwd_approx}
    U(\mathbf{\overline{X}}[L]) \approx \sum_{n=0}^{L} \gamma^n r[n].
\end{equation}
In Appendix \ref{subsec:proof_per_ts_rwd}, we prove that by defining the per-timestep reward as $r[n] = \sum_{j=1}^{N} r_j[n] \ \forall \ n > 0$ and $r[0] = \sum_{j=1}^{N} \log(\overline{X}_j[0])$ where 
\begin{equation} \label{eq:per_ts_rwd}
    r_j[n] = \log\Bigg((1-1/B)\Big(1 + \frac{R_j[n]}{(B-1)\overline{X}_j[n-1]}\Big)\Bigg),
\end{equation}
and setting the discount factor $\gamma \rightarrow 1$, we satisfy the approximation postulated in \eqref{eq:rwd_approx}. In case all BSs choose to remain off, we apply a large negative reward (empirically set to $-\kappa N$ for some constant $\kappa$) for that time-step during training. This does not impact \eqref{eq:rwd_approx} as no optimal joint policy $\pi^*$ would have all the BSs remain off in any time slot. Finally, we define the observation $o_i$ received in each timestep $n$ by BS $i$ as
\begin{equation}
    o_i[n] = \langle \overline{X}_i[n-1],S_i[n-1],I_i[n-1] \rangle,
\end{equation}
where $n-1$ denotes that the observation seen at the beginning of the $n^{\mathrm{th}}$ time slot by the BS is the average rate, signal and interference power of the UE it serves from the previous data frame (enforcing causality). Both $S_i$ and $I_i$ can be obtained in practice from UE feedback comprising of CSI, RSRP and RSRQ (Reference Signal Received Power and Quality) measurements (refer \cite{3gpp.38.214}, \cite{dahlman20205g} for details). Note that the signal and interference terms $S_i$ and $I_i$ act as partial observations of $\overline{X}_j \ \forall \ j \neq i$ and $\{g_{ij}\}$. This can be seen from \eqref{eq:rate_update} where knowledge of $S_i[n-1]$ and $I_i[n-1]$ provides the BS a better estimate of $\{g_{ij}[n-1]\}$, which is highly correlated with $g_{ij}[n]$ in a slow fading channel model. Moreover, $S_i[n-1] = g_{ii}[n-1] a_i[n-1]$, hence $o_i[n]$ also provides the action of BS $i$ in the previous time step $n-1$ as an input while choosing its action for time step $n$.

We will now describe a distributed RL framework that can be adapted to this \textit{DEC-POMDP} formulation, while catering to a contention-based medium access mechanism and providing for partial observability of the system state.

\section{Adapting Independent DQN to a Medium Access DEC-POMDP} \label{sec:ind_dqn}
The \textit{DEC-POMDP} formulation presented in Section \ref{subsec:med_acc_dec_pomdp} essentially entails each BS $i$, in a time slot $n$, observing $o_i[n]$, taking an action $a_i[n]$, and receiving a common reward $r[n]$ once all BSs in the layout have taken an action. This is summarized in the 1 state MDP depicted on the left in Fig. \ref{fig:MDP}, with the term inside the square denoting the observation $o_i$ of BS $i$. Note that while the word MDP may be a misnomer here, it is only used to capture the nature of the transition diagram and does not confer a MDP's mathematical properties\footnote{We are dealing with a DEC-POMDP, hence a policy $\pi_i$ at BS $i$ is a function that takes as input the action-observation history $\vec{o}_i$ and outputs $a_i$. This will be provided for in Section \ref{subsec:drqn} using recurrency.\label{footnote:dec_pomdp}}. However, this formulation fails to exploit a key aspect of medium access: the asynchronous nature of contention-based access in LBT based spectrum sharing mechanisms. Essentially, when a BS performs CCA, we assumed that a subset of the BSs that have already chosen to transmit will be transmitting a unique preamble (akin to a RTS frame)\footnote{Receipt of the confirmatory CTS message at the BS is not required to make a decision, thus eliminating the associated overhead and latency.}. This frame would contain the MAC address of the transmitter, enabling a BS listening in on this transmission to map the received signal to the corresponding transmitter. Such a strategy could be easily supported in LAA/NR-U, such that the energy measured at a BS could not only be apportioned into the energy from each BS that is already transmitting but also mapped to them.

\subsection{EOS and CON States} \label{subsec:eos_con}
\begin{figure*}[!ht]
    \centering
    \includegraphics[width =  6in]{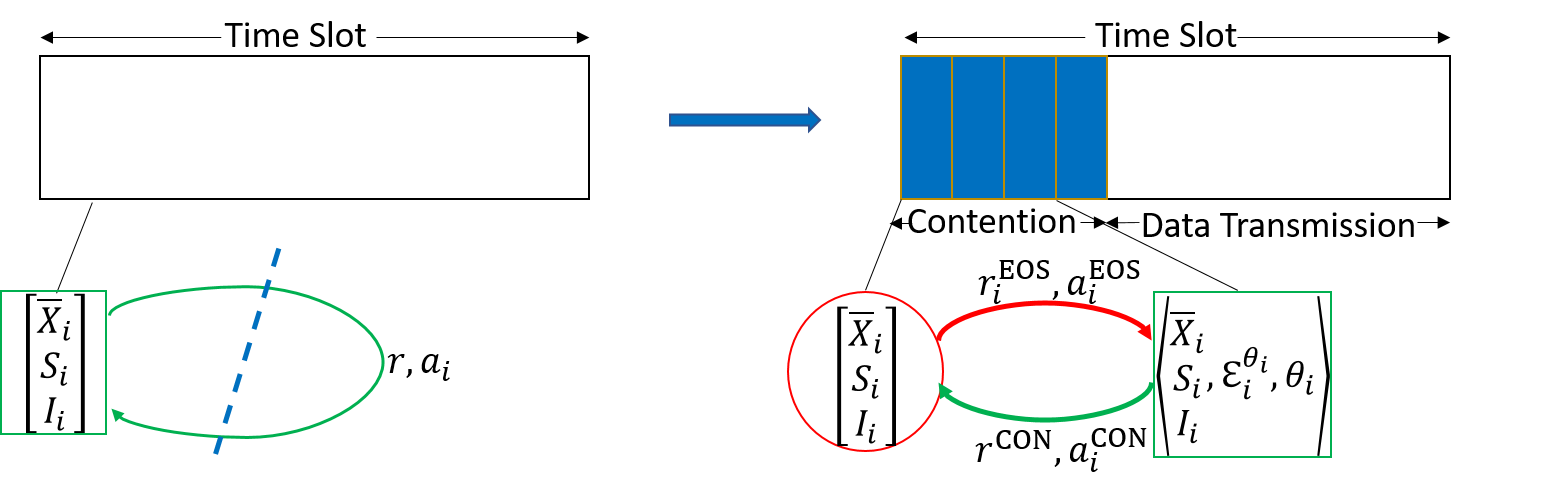}
    \caption{The 2 state MDP at each agent capturing the actions taken and reward obtained on transitioning between the End-Of-Slot (EOS) and Contention (CON) states}
    \label{fig:MDP}
\end{figure*}

This motivates us to consider the partitioning of the 1-state MDP into 2 states, End-Of-Slot (\textcolor{red}{EOS}) and Contention (\textcolor{green}{CON}), as shown on the right in Fig. \ref{fig:MDP}. In each time slot $n$, a back-off counter $\theta_i \in \{0, \ldots, \mathrm{CW}-1\}$ is randomly generated for each BS $i$ from a fixed contention window (CW). When that counter expires, the BS measures the energy from each already transmitting BS, denoted by the $N$-dimensional vector $\mathcal{E}_i^{\theta_i} = \{\mathcal{E}_{ij}^{\theta_{i}}\}_{j\in[N]}$, such that $\mathcal{E}_{ij}^{\theta_{i}}$ is the energy received at BS $i$ due to an ongoing transmission between BS $j$ and UE $j$, and is given by
\begin{equation} 
    y_{ij}^{\theta_{i}} = \ \sqrt{P_t}h_{ij}'a_j\mathbf{1}_{\theta_j < \theta_i} + z_{ij} \;\;\;\;\;\;\;\; \mathcal{E}_{ij}^{\theta_{i}} = \big|y_{ij}^{\theta_{i}}\big|^2,
    \label{eq:rec_signal_BS}
\end{equation}
where $z_{ij} \sim \mathcal{CN}(0,\sigma^2_{\mathrm{BS}})$, $\mathbf{1}_{\theta_j < \theta_i} = 1$ if $\theta_j < \theta_i$ and $y_{ij}^{\theta_{i}}$ is the received signal corresponding to $\mathcal{E}_{ij}^{\theta_{i}}$. In other words, $\mathcal{E}_{ij}^{\theta_{i}}$ will have a component other than noise only if BS $j$ is placed before BS $i$ in the contention queue and has chosen to transmit. Note that if $\theta_j > \theta_i$, then the value of $a_j$ is inconsequential in \eqref{eq:rec_signal_BS}, ensuring the construction of $\mathcal{E}_i^{\theta_i}$ is causal. It also shows that BS $i$ has only partial observability of the action of the other BSs. For instance, if $a_j = 0 \ \forall j$, then BS $i$ cannot deduce which entries of $\mathcal{E}_{ij}^{\theta_{i}}$ correspond to BSs whose counter $\theta_j > \theta_i$ versus BSs whose counter $\theta_j < \theta_i$ and have chosen $a_j = 0$. However, this partial observability can be remedied to some extent at BS $i$ by knowledge of its own counter $\theta_i$. For instance, if $\theta_i$ is large and $a_j = 0 \ \forall j$, then BS $i$ would be privy to the fact that most BSs $j$, $j \neq i$, have chosen not to transmit. Consequently, $\mathcal{E}_{i}^{\theta_{i}}$ and $\theta_i$ are defined to be part of the observation in the CON state.

Moreover, we note that $h'_{ii} = 0$ . It may not be immediately apparent why one would need to retain $\mathcal{E}_{ii}^{\theta_{i}}$ as part of the state definition; experimental evidence and a possible explanation is presented in Section \ref{subsec:energy_vector_length}. 

BS $i$ then selects an action $a_i^{\mathrm{CON}}$ based on the policy $\pi_i^{\mathrm{CON}}: \langle \vec{o}_i, \vec{\mathcal{E}}_i^{\theta_i}, \vec{\theta}_i \rangle \rightarrow A_i$ (refer footnote \ref{footnote:dec_pomdp}). If a BS decides to transmit, data transmission takes place until the end of slot. At the end of slot $n$, using the action vector $\mathbf{a}[n]$, average rates $\mathbf{\overline{X}}[n-1]$ and path gain matrix $\mathbf{G}[n]$, we compute $r^{\mathrm{CON}}[n] = r[n]$ using \eqref{eq:per_ts_rwd}. The average rates are updated to $\mathbf{\overline{X}}[n]$ using \eqref{eq:rate_update}, and along with the signal power $S_i[n]$ and interference power $I_i[n]$ measured at each UE $i$, are used to calculate the observation seen by each BS in the next time slot, given by $o_i[n+1] = \langle \overline{X}_i[n], S_i[n], I_i[n] \rangle$. Hence we denote the EOS observation by $\mathbf{o}^{\mathrm{EOS}}_i = o_i$, and the CON observation by $\mathbf{o}^{\mathrm{CON}}_i = \langle o_i, \mathcal{E}_i^{\theta_i}, \theta_i \rangle$. Both $r^{\mathrm{EOS}}_i$ and $a^{\mathrm{EOS}}_i$ default to 0, since these correspond to a transition within the time slot.

\subsection{Deep Q-Networks and Independent DQN}\label{subsec:dqn}
In a single agent, fully-observable RL setting, an agent observes the current state $s[n]$, chooses an action $a[n]$ according to a policy $\pi$, receives a reward $r[n]$ and transitions to a new state $s[n+1]$. The objective is to learn the optimal policy $\pi^*$ that maximizes the expected discounted sum of rewards $\sum_{n=0}^{\infty} \gamma^n r[n]$. Denote by $\mathcal{Q}_{\pi}(s[n],a)$ the expected discounted reward earned by the agent starting from state $s[n]$, taking action $a$, and thereafter following $\pi$. Hence the $\mathcal{Q}$-value corresponding to the optimal policy $\pi^*$
\begin{equation}
    \mathcal{Q}^*(s[n],a) = \max_{\pi}\mathcal{Q}_{\pi}(s[n],a)
\end{equation}
is given by the recursive Bellman optimality equation with $n' = n+1$ \cite{sutton2018reinforcement} 
\begin{equation} \label{eq:Q_gen_update}
    \mathcal{Q}^*(s[n],a) = \mathbb{E}\Big[ r[n'] + \gamma\max_{a'}{Q}^*(s[n'],a') \Big| s[n],a\Big].
\end{equation}

In deep $\mathcal{Q}$-learning, $\mathcal{Q}_{\pi}(s,a)$ is represented by a neural network whose weights are optimized by minimizing $\mathbb{E}_{s,a,r,s'}[(y^{\mathrm{DQN}} - Q(s,a))^2]$ at each iteration, with $y^{\mathrm{DQN}} = r + \gamma \max_{a'} \mathcal{Q}^{-}(s',a')$. Here $\mathcal{Q}^{-}(s,a)$ denotes a target $\mathcal{Q}$ network that is frozen for a few iterations while updating the online network $\mathcal{Q}(s,a)$.

DQN has been extended to cooperative multi-agent settings, in which each agent $i$ observes the complete state $s$, selects an individual action $a_i$, and receives a team reward $r$, shared among all agents. In \cite{tampuu2017multiagent}, they address this setting with a framework that combines DQN with independent $\mathcal{Q}$-learning, such that each agent $i$ simultaneously learns its own $\mathcal{Q}$ function $\mathcal{Q}_i(s,a_i)$. While Independent Q-learning can in principle lead to convergence problems due to each agent seeing a non-stationary environment, it has been surprisingly successful empirically.

\subsection{Deep Recurrent Q-Learning} \label{subsec:drqn}
In Section \ref{subsec:dec_pomdp}, we defined a BS's policy $\pi_i$ as a mapping from the local action-observation history $\vec{\omega}_i$ to the action space $A_i$. One way to incorporate the action-observation history into Deep Q-Learning is to stack the observations obtained over a finite number of time steps and feed all these observations into the DQN. However, in \cite{hausknecht2015deep}, they hypothesize that a Deep Recurrent Q-Network, which is a combination of DQN with a Long Short Term Memory (LSTM) \cite{hochreiter1997long} layer can better approximate actual Q-values from sequences of observations, leading to better policies in partially observed environments. 

One example of a DQN architecture incorporating an LSTM layer is depicted in Fig. \ref{fig:dqn_arch}. In time slot $n$, the DQN receives as input the hidden state $state\_h_{i,n-1}$ and cell state $state\_c_{i,n-1}$ of the LSTM layer in the DQN from the last time slot. Each DQN then outputs the updated $state\_h_{i,n}$ and $state\_c_{i,n}$ which is fed into the DQN in the next time step. Intuitively, the hidden and cell state are compressed representations of the local action-observation history $\vec{o}_i$, such that the policy learnt by the DQN is indeed a mapping from $\vec{\omega}_i$ to $A_i$. 

\subsection{Bellman Updates for EOS and CON States} \label{subsec:bellman_q_val}
Solving the medium access \textit{DEC-POMDP} entails finding the optimal joint policy $\mathbf{\pi}^*$ that maximizes $\sum_{n=0}^{L} \gamma^n r[n]$. This requires computing the optimal policy $\pi_i^*$ at each BS $i$. To this end, we will adopt the Independent DQN framework and define two $\mathcal{Q}$ networks at each BS $i$, $\mathcal{Q}_i^{\mathrm{EOS}}$ and $\mathcal{Q}_i^{\mathrm{CON}}$. As a first step towards addressing partial observability, each of these $\mathcal{Q}_i$ networks will be modelled as recurrent $\mathcal{Q}$-networks as described in Section \ref{subsec:drqn}. However, in addition to partial state observability, a decentralized setting makes the action $a_j$ of BS $j$ not observable at BS $i$ for $i \neq j$. In other words, each BS $i$ would not be able to observe the entire $\mathbf{a}-\{a_i\}$ vector. This is where the energy vector $\mathcal{E}_i^{\theta_i}$ comes in handy. As described in Section \ref{subsec:eos_con}, it enables BS $i$ to observe the action of BS $j$ if $\theta_j < \theta_i$ and $a_j = 1$.

This indirect exchange of ``messages" between BSs via $\mathcal{E}_i^{\theta_i}$ during contention has parallels to the technique for inter-agent communication in a \textit{DEC-POMDP} via a limited-bandwidth channel \cite{foerster2016learning}. A technique called \textit{differentiable inter-agent learning} (DIAL) is proposed in \cite{foerster2016learning} that not only uses deep Q-learning \cite{mnih2015human} with a recurrent network to address partial observability \cite{hausknecht2015deep}, but also provides for the passage of customized real-valued messages between agents, such that gradients can be pushed through the communication channel yielding a system that is end-to-end trainable across agents. In \cite{foerster2016learning}, they employ direct connections between the output of one agents network and the input of another, via a \textit{discrete/regularize unit} (DRU), allowing the message to be learnt during training. However, in our case, the contents of the ``message" $\mathcal{E}_i^{\theta_i}$ cannot be modified to improve learning at the BS. In Section \ref{sec:Results}, we will provide experimental evidence that the vector $\mathcal{E}_i^{\theta_i}$ nevertheless enables the learning of a competitive decentralized medium access policy $\pi$.

\begin{figure}[!ht]
    \centering
    \includegraphics[width=3.1in]{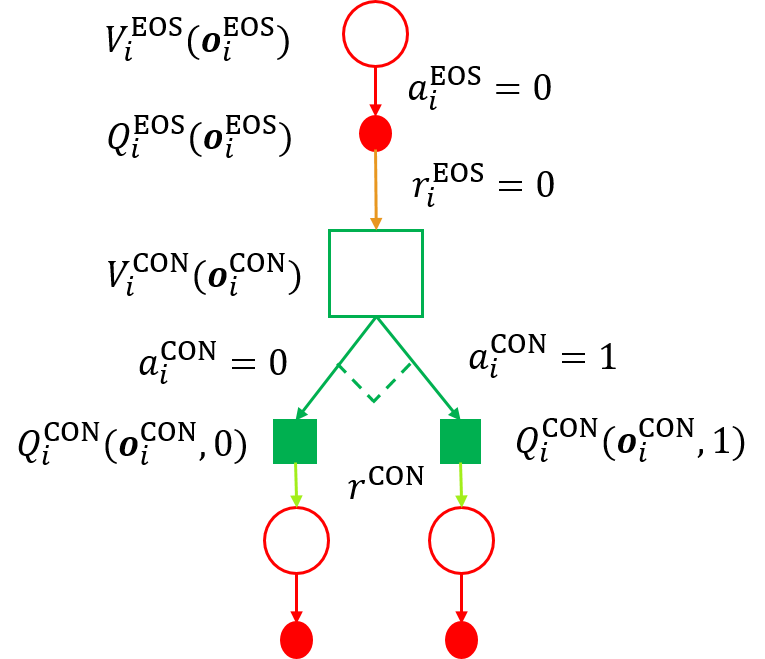}
    \caption{Backup diagram for calculating $\mathcal{Q}$ values at both $\mathrm{EOS}$ and $\mathrm{CON}$. Each open circle denotes a $\mathrm{EOS}$ state, and each open square a $\mathrm{CON}$ state. Each solid circle denotes a $\mathrm{EOS}$ state-action pair ($a_i^{\mathrm{EOS}}=0$), and each solid square a $\mathrm{CON}$ state-action pair. The dotted arc represents that the maximum of the two branches is taken.}
    \label{fig:backup_diagram}
\end{figure}

Hence, using \eqref{eq:Q_gen_update}, in accordance with the backup diagram depicted in Fig. \ref{fig:backup_diagram}, we can derive the sampled $\mathcal{Q}$-value updates for $\mathbf{o}^{\mathrm{EOS}}_i$ and $\mathbf{o}^{\mathrm{CON}}_i$ as 
\begin{align}
     &\mathcal{Q}_i^{\mathrm{EOS}}(\mathbf{o}_i^{\mathrm{EOS}}) = \gamma \max_{a_i^{\mathrm{CON}}} \mathcal{Q}_i^{\mathrm{CON}}(\mathbf{o}_i^{\mathrm{CON}},a_i^{\mathrm{CON}}) \label{eq:q_eos}\\
    &\mathcal{Q}_i^{\mathrm{CON}}(\mathbf{o}_i^{\mathrm{CON}},a_i^{\mathrm{CON}}) = r^{\mathrm{CON}} + \gamma \mathcal{Q}_i^{\mathrm{EOS}}(\mathbf{o}_i^{\mathrm{EOS}}), \label{eq:q_con}
\end{align}
where we utilize $r^{\mathrm{EOS}}_i = a^{\mathrm{EOS}}_i = 0$, hence $\mathcal{Q}_i^{\mathrm{EOS}}(\mathbf{o}_i^{\mathrm{EOS}}) = \mathcal{Q}_i^{\mathrm{EOS}}(\mathbf{o}_i^{\mathrm{EOS}},0)$. Also note that  $r^{\mathrm{CON}} = r$ and $a^{\mathrm{CON}}_i = a_i$. We replaced $\mathbf{s}$ with $\mathbf{o}_i^{\mathrm{EOS}}$ and $\mathbf{o}_i^{\mathrm{CON}}$ in \eqref{eq:q_eos} and \eqref{eq:q_con} respectively,
since $\mathcal{Q}_i^{\mathrm{EOS}}$ and $\mathcal{Q}_i^{\mathrm{CON}}$ are recurrent $\mathcal{Q}$-networks. The recurrency implicitly captures the fact that the observation-action history is also part of the input to each $\mathcal{Q}$ network.

An advantage of these alternating $\mathcal{Q}$-updates is we eliminate the need for double deep $\mathcal{Q}$-learning using a target $\mathcal{Q}$-network as described in Section \ref{subsec:dqn}. Since a typical Bellman equation would contain $\mathcal{Q}$ values corresponding to the same state space on both sides of the equation, double deep $\mathcal{Q}$-learning is introduced to avoid \textit{maximization bias} \cite{sutton2018reinforcement}. In \eqref{eq:q_eos} and \eqref{eq:q_con}, we have $\mathcal{Q}$ values corresponding to different state spaces on both sides, hence circumventing the issue. At the same time, it must be noted that \eqref{eq:q_eos} and \eqref{eq:q_con} are equivalent to the original Bellman equation, albeit with a discount factor of $\gamma^2$. This can be seen by substituting \eqref{eq:q_eos} in \eqref{eq:q_con}. We now describe combining recurrent DQNs with message-passing between BSs to generate an episode using the example depicted in Fig. \ref{fig:pic_algo}. 

\subsection{Generating an episode using DQN's} \label{subsec:gen_episode}
\begin{figure*}
    \centering
    \includegraphics[width = 7in]{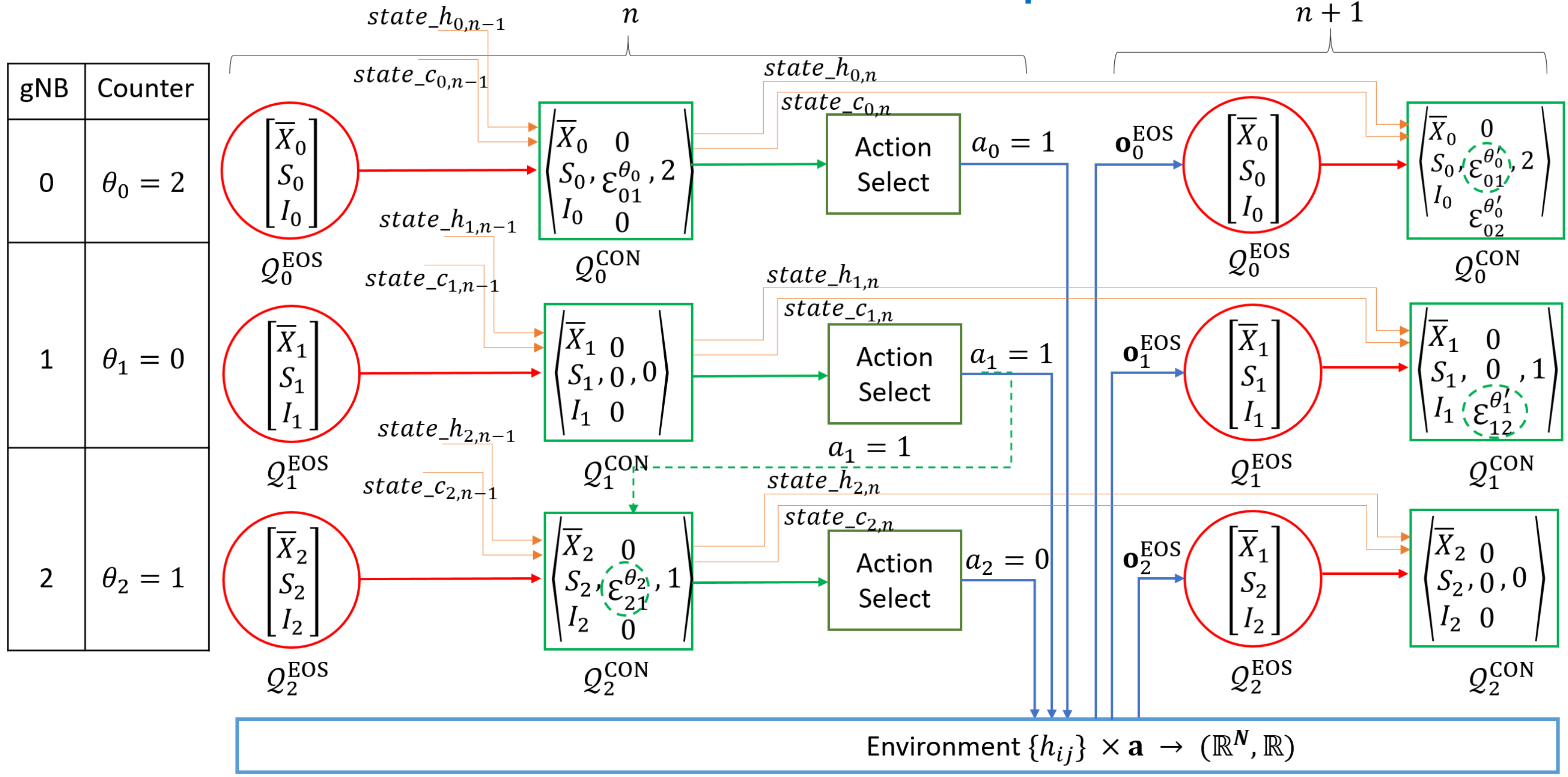}
    \caption{Information flow between BS's and the environment. The evolution of the hidden and cell state, $state\_h_{i,n}$ and $state\_c_{i,n}$ is shown only for $\mathcal{Q}^{\mathrm{CON}}$, but holds for $\mathcal{Q}^{\mathrm{EOS}}$ too. The red circle denotes the EOS DQN and the green square, the CON DQN.}
    \label{fig:pic_algo}
\end{figure*}

An episode refers to a collection of $L$ consecutive time slots. Consider the $n^{\mathrm{th}}$ time slot in the episode. At the beginning of time slot $n$, a random counter $\theta_i$ is drawn for each BS $i$ as shown in the table on the left in Fig. \ref{fig:pic_algo}. Note that for time slot 0, the same observation $<\overline{X}[0],0,0>$ with $\overline{X}[0] = 10^{-2}$ is fed to $\mathcal{Q}^{\mathrm{EOS}}_i$ at each BS $i$. In subsequent time steps, $\mathbf{o}^{\mathrm{EOS}}_i$ is computed in the previous time slot as explained in Section \ref{subsec:eos_con}. Recall that $\mathcal{Q}_i^{\mathrm{EOS}}(\mathbf{o}_i^{\mathrm{EOS}}) = \mathcal{Q}_i^{\mathrm{EOS}}(\mathbf{o}_i^{\mathrm{EOS}},0)$. On the other hand, each CON DQN $\mathcal{Q}^{\mathrm{CON}}_i$ outputs two $\mathcal{Q}$-values corresponding to the actions 0 and 1, with 
\begin{equation} \label{eq:action_policy}
    a_i = \pi^{\mathrm{CON}}_i (\mathbf{o}^{\mathrm{CON}}_i) = \underset{a \in A_i} {\mathrm{arg\ max\ }} \mathcal{Q}^{\mathrm{CON}}_i (\mathbf{o}^{\mathrm{CON}}_i,a).
\end{equation}
Note that if we were generating an episode during the training of the algorithm, we would use an $\epsilon$-greedy policy to choose $a_i$ to provide for the exploration-exploitation trade-off \cite{mnih2015human}. Hence, w.p. $1-\epsilon$ we would choose $a_i$ using \eqref{eq:action_policy}, and w.p. $\epsilon$ we would pick $a_i$ randomly.

In the $N=3$ example shown in Fig. \ref{fig:pic_algo}, since $\theta_1 = 0$, BS 1 goes first and measures the energy from ongoing transmissions to compute $\mathcal{E}_1^{\theta_1}$ using \eqref{eq:rec_signal_BS}. It senses no other BS's transmitting ($\mathcal{E}^{\theta_1}_{1} = [0,0,0]$, ignoring the noise term in \eqref{eq:rec_signal_BS} for simplicity), and in combination with the average rate $\overline{X}_1[n-1]$, signal $S_1[n-1]$ and interference power $I_1[n-1]$ of the UE it serves from the previous time slot, it determines $a_1[n]$ using the policy given in \eqref{eq:action_policy}. Let us assume it chooses to transmit (note that transmission is not a given simply because $\mathcal{E}^{\theta_1}_{1} = [0,0,0]$, it is a complex decision made by the DQN based on many factors). BS 2 is scheduled next and it detects BS 1 is transmitting such that $\mathcal{E}^{\theta_2}_{21}$ is non-zero and $\mathcal{Q}^{\mathrm{CON}}_2$ instructs it not to transmit. Finally BS 0 also detects a non-zero $\mathcal{E}^{\theta_0}_{01}$, but chooses to transmit. At this point, we save the tuple $\langle \mathbf{o}^{\mathrm{EOS}}_i,\mathbf{o}^{\mathrm{CON}}_i \rangle$ at each BS $i$ . Note that while the training procedure, elaborated in Section \ref{subsec:training_proc}, will require training both the CON and EOS DQN, testing the learnt policy requires only the CON DQN as is evident from \eqref{eq:action_policy}.

Once all the BS's have taken an action $a_i$, the action vector $\mathbf{a}$ in combination with the channel gains $h_{ij}$ are used to calculate $r[n] \in \mathbb{R}$ and the updated average rates $\mathbf{\overline{X}}[n] \in \mathbb{R}^N$. These determine the observations $\mathbf{o}^{\mathrm{EOS}}_i$ for the next time slot. At this point, we save the quadruple $\langle \mathbf{o}^{\mathrm{CON}}_i,a_i,r^{\mathrm{CON}},\mathbf{o}^{\mathrm{EOS}}_i \rangle$ at each BS $i$. In the next time slot $n+1$, a new counter $\theta'_i$ is drawn at each BS $i$ and the process is repeated.

\section{Simulation Details} \label{sec:sim_details}
The performance metric is the expected cumulative reward $\sum_{n=0}^{L} \gamma^n r[n]$ introduced in Section \ref{subsec:med_acc_dec_pomdp}, with $r[n]$ given by \eqref{eq:per_ts_rwd}. A smoothing window of $B=10$, $\gamma = 1 - 10^{-6}$ and $L = 2000$ time slots are used in all the subsequent simulations.

\subsection{Data Generation \& Pre-processing}
We consider an indoor hotspot deployment, a scenario that is intended to capture typical indoor situations such as office environments comprised of open cubicle areas, walled offices, open areas and corridors (InH-Office in 3GPP TR 38.901 \cite{3gpp.38.901}). The BSs are located at 10, 30, 50, 70, 90, 110 meters on the x-axis, 15 and 35m on the y-axis and are mounted at a height of 3m on the ceiling. 120 UEs are uniformly distributed in a 120m x 50m layout and have a height of 1.5m. Each BS is associated with 10 UEs. The pathloss (PL) model between nodes(BS and UEs) captures Line-Of-Sight (LOS)/ Non-Line-Of-Sight (NLOS) properties of a link, frequency dependent path loss for LOS/NLOS links and shadowing as part of large-scale fading parameters, and is given in Table \ref{tab:pathloss}. In accordance with the indoor - open office model \cite{3gpp.38.901}, the links are designated as LOS/ NLOS probabilistically with $\mathrm{Pr_{LOS}}$ given by
\begin{align}
    \mathrm{Pr_{LOS}} = \begin{cases}1 & d_{\mathrm{2D}} \leq 5\mathrm{m}\\
    \exp\Big(-\frac{d_{\mathrm{2D}}-5}{70.8}\Big) & 5\mathrm{m} < d_{\mathrm{2D}} \leq 49\mathrm{m}\\
    0.54\exp\Big(-\frac{d_{\mathrm{2D}}-49}{211.7}\Big) & 49\mathrm{m} < d_{\mathrm{2D}.}
    \end{cases}
\end{align}
with $d_{\mathrm{2D}}$ denoting the distance between the BS and UE on the floor, while $d_{\mathrm{3D}}$ will be used to denote the distance between the highest points of the BS and UE respectively.

The center frequency $f_c$ used for modeling is 6 GHz. From these 12 BSs, various configurations of $N=4$ BSs and its associated UEs are chosen for the simulations, for e.g. refer Fig. \ref{fig:layout_1} and Fig. \ref{fig:layout_2}.
\begin{table*}
    \centering
    \includegraphics[width=5in]{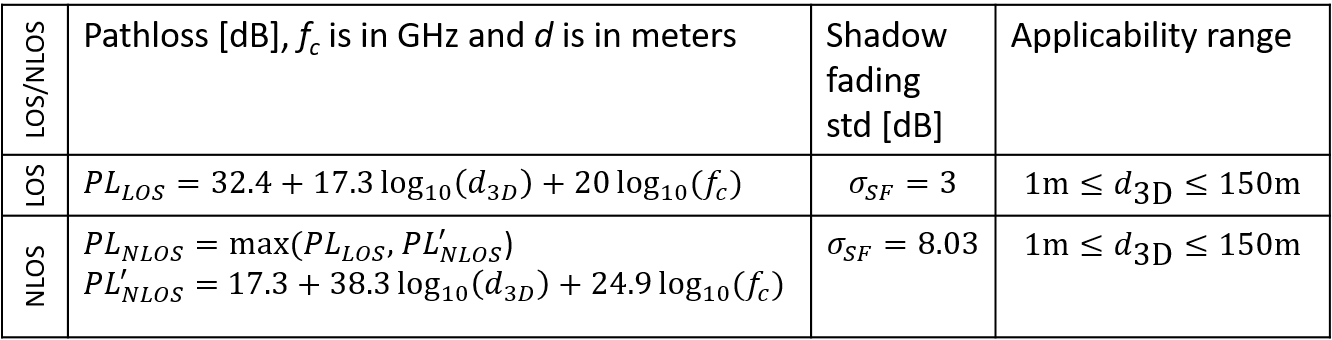}
    \caption{InH-Office PL Model. Shadow fading distribution is log-normal. Reproduced from \cite{3gpp.38.901}.}
    \label{tab:pathloss}
\end{table*}

At each BS, to select the active UE, we have 10 UEs to choose from. From these $10^4$ possible configurations of 4 UEs, $9^4$ are chosen for training the RL algorithm, and the rest are used for randomly sampling validation and test configurations. Replay memories $D_{\mathrm{EOS}}$ and $D_{\mathrm{CON}}$, modelled as double ended queues, are created, each containing $L_D = 9^4$ episodes, one for each possible training configuration. Each episode in the initial replay memory is generated with an $\epsilon=1$ greedy policy, i.e. at each time step in the episode, $a_i$ is chosen randomly at each BS $i$. However, $\epsilon$ decays over the course of training for generating new episodes as is described in Section \ref{subsec:training_proc}.

Within each episode, slow fading with a first order IIR filter is used to model the time evolution of each of the channel coefficients $h_{ij}$ and $h'_{ij}$. Denote the large scale fading (given by the InH-Office PL model) path gain coefficient as $g_{\mathrm{ls}} = g_{0}$, and the small scale fading coefficient as $h_{\mathrm{ss}}$. Then we have
\begin{align}
    &h_{\mathrm{ss}}[n] = (1-\alpha) h_{\mathrm{ss}}[n-1] + \alpha z[n]  \\
    &g[n] = g_0\big|h_{\mathrm{ss}}[n]\big|^2, 
    \label{eq:fading}
\end{align}
with $h_{\mathrm{ss}}[0] = 1$, $z[n]\sim \mathcal{CN}(0,\sigma^2_{\mathrm{ss}})$ and $\sigma^2_{\mathrm{ss}} = (1 - (1-\alpha)^2)/\alpha^2$. The length of one slot $T$ is given by the COT which ranges from 1 to 9 ms. Solving $0.5 = (1 - \alpha)^n$ with $\alpha = 0.01$ yields $n = 69$.  Hence, the time taken for the channel to decorrelate $50\% = 69T$, which is on the order of 100 ms, typical for a pedestrian setting. Key parameters used for generating the data are summarized in Table \ref{tab:sim_param}a.

Both $S_i$ and $I_i$ in the definition of $\mathbf{o}^{\mathrm{EOS}}_i$ are normalized by the standard deviation of the BS-UE path gains $\sigma_{gU}$ before being input to the CON and EOS DQNs. Similarly, each entry of $\mathcal{E}^{\theta_i}_{i}$ in $\mathbf{o}^{\mathrm{CON}}_i$ is  normalized by the standard deviation of the BS-BS path gains $\sigma_{gg}$. 

\subsection{DQN Architecture}
\begin{figure}
    \centering
    \includegraphics[width=3.4in]{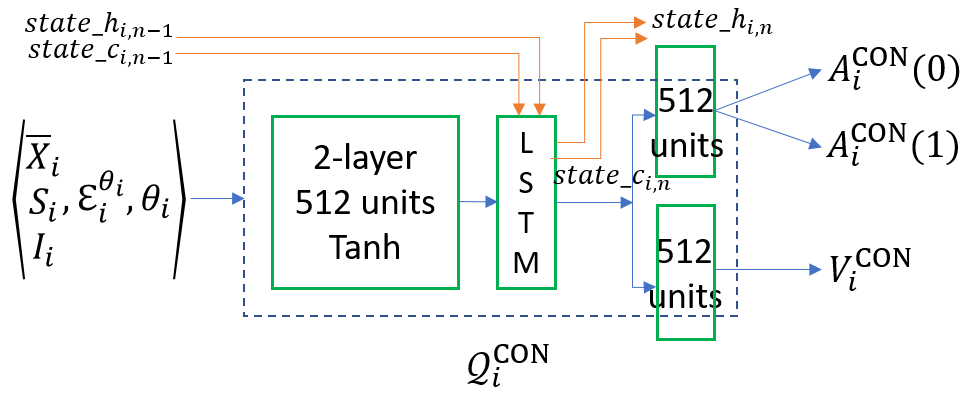}
    \caption{NN Architecture of $\mathcal{Q}_i^{\mathrm{CON}}$, depicting the passage of the hidden and cell state of the LSTM operation from one time slot to the next.}
    \label{fig:dqn_arch}
\end{figure}

Both the CON and EOS DQN's have a similar architecture, the only difference being the size of the input. The neural network (NN) architecture of a $\mathcal{Q}_i^{\mathrm{CON}}$ DQN is depicted in Fig. \ref{fig:dqn_arch}. Since the input matrix to an LSTM\footnote{https://pytorch.org/docs/stable/generated/torch.nn.LSTM.html} has three dimensions $(\mathrm{batch},\mathrm{seq\_len},\mathrm{input\_size})$, with $\mathrm{seq\_len}$ capturing the temporal dependence, while the input to a fully connected layer is two dimensional, with the second dimension being $\mathrm{input\_size}$, we flatten the $\mathrm{seq\_len}$ dimension such that the input to the DQN is of size $(\mathrm{batch} \times \mathrm{seq\_len},\mathrm{input\_size})$. 

The input state is fed into a 2-layer fully connected DNN, 512 neurons per layer, and tanh activation. Hence these two layers are applied per time-step separately. At this stage, the output of size $(\mathrm{batch}\times\mathrm{seq\_len},512)$ is reshaped into $(\mathrm{batch},\mathrm{seq\_len},512)$ before being fed into a LSTM layer, having a hidden state $state\_h_{i,n}$ and cell state $state\_c_{i,n}$ of size 256.  Only the value of the last time step (corresponding to index $\mathrm{seq\_len}-1$) is extracted from the LSTM output, such that the output is of size $(\mathrm{batch},256)$. The LSTM output is then fed into two fully connected layers, one of which outputs a $V$ value, while the other outputs 2 $A$ values corresponding to both possible actions. The advantage $A(s,a)$ is the gain, in expected cumulative reward $Q(s,a) - V(s)$, of choosing action $a$. The Q value is then computed as 
\begin{equation}
    Q_i(s,a) = V_i(s) - \frac{1}{|A_i|}A_i(s,a).
\end{equation}
This approach is known as \textit{dueling} \cite{wang2016dueling}, and helps in learning the state-value function efficiently. Finally, $Q^{\mathrm{EOS}}$ always has a default action of $a^{\mathrm{EOS}} = 0$, so the Q-value of $a^{\mathrm{EOS}} = 1$ is discarded.

\subsection{DQN Training Procedure} \label{subsec:training_proc}
In each training iteration, an episode is generated as described in Section \ref{subsec:gen_episode} using an $\epsilon$-greedy policy at each BS with $\epsilon$ decaying from 1 to 0.25 uniformly with each iteration. An episode yields $L$ $\langle \mathbf{o}^{\mathrm{EOS}}_i,\mathbf{o}^{\mathrm{CON}}_i \rangle$ tuples and $\langle \mathbf{o}^{\mathrm{CON}}_i,a_i,r^{\mathrm{CON}},\mathbf{o}^{\mathrm{EOS}}_i \rangle$ quadruples from each BS $i$, denoted as $E_{\mathrm{EOS},i}$ and $E_{\mathrm{CON},i}$ respectively. These episodes are collectively appended as a new row in replay memories $D_{\mathrm{EOS}}$ and $D_{\mathrm{CON}}$ respectively.

Then $\mathrm{batch} = 5000$ episodes are randomly sampled from both memories, and starting from a random point in each episode, $\mathrm{seq\_len} = 50$ consecutive transitions are chosen for training. We employ $\mathrm{seq\_len} = 50$ to carry the LSTM hidden state forward to some extent during training, while also better adhering to a DQN's random sampling policy for training than $\mathrm{seq\_len} = L$ (refer to the discussion on Bootstrapped Random Updates in \cite{hausknecht2015deep}). Subsequently, we employ the 2-stage Bellman update equations \eqref{eq:l_eos} and \eqref{eq:l_con} derived in Section \ref{subsec:bellman_q_val} to calculate $\mathrm{batch}$ labels $L^{\mathrm{EOS}}_i$ and $L^{\mathrm{CON}}_i$ corresponding to the predictions of  $\mathcal{Q}^{\mathrm{EOS}}_i$ and $\mathcal{Q}^{\mathrm{CON}}_i$ respectively at each BS $i$ as follows
\begin{align}
     L_i^{\mathrm{EOS}}(\mathbf{o}_i^{\mathrm{EOS}}) = \gamma& \max_{a_i^{\mathrm{CON}}} \mathcal{Q}_i^{\mathrm{CON}}(\mathbf{o}_i^{\mathrm{CON}},a_i^{\mathrm{CON}}) \label{eq:l_eos}\\
     L_i^{\mathrm{CON}}(\mathbf{o}_i^{\mathrm{CON}},a_i^{\mathrm{CON}})& = r^{\mathrm{CON}} + \gamma \mathcal{Q}_i^{\mathrm{EOS}}(\mathbf{o}_i^{\mathrm{EOS}}). \label{eq:l_con}
\end{align}
We then update the weights of each DQN using the mean squared error between the prediction and label as the loss function. The remaining parameters used for training the NNs are summarized in Table \ref{tab:sim_param}b. The choice of initial learning rate $\eta$ depends on the layout we are training on, and is given for both layouts utilized in Section \ref{sec:Results}. It should be noted that the outlined training procedure is performed completely offline prior to deployment, since it requires centralized computation of the reward $r^{\mathrm{CON}}$. However, it can be adapted to an online setting as well (refer Section \ref{subsec:online}).
\begin{table*}
	\caption[Simulation Parameters] {Simulation Parameters}
	\label{tab:sim_param}
	\centering
	\subfloat[Data Generation Parameters]{\begin{tabular}{ |p{3.5cm}|p{2cm}|}
		\hline
		$N$ & 4\\\hline
		Transmit Power $P_t$ & 23 dBm \\\hline
		Noise PSD & -174 dBm/Hz \\\hline
		Bandwidth & 20 MHz \\\hline
		UE Noise Figure & 9 dB\\\hline
		BS Noise Figure & 5 dB\\\hline
		Fading Coefficient $\alpha$ & 0.01\\\hline
	\end{tabular}\label{tab:dat_gen_param}}
	\hspace{0.2in}
	\subfloat[DQN Training Parameters]{
	\begin{tabular}{ |p{3.5cm}|p{3.5cm}|}
		\hline
		Initial Learning Rate $\eta$ & L1: $2\times10^{-5}$ L2:  $10^{-4}$\\\hline
		Learning Rate Decay &  0.85 ~/~ 500 updates\\\hline
		Weight Decay & 0.001 \\\hline
		Optimizer & Adam\cite{kingma2014adam}\\\hline
		Batch Size & 5000\\\hline
		Training Iterations & 15000\\\hline
	\end{tabular}\label{tab:dqn_training_param}}
\end{table*}

\subsection{PF Scheduler and ED Threshold Baselines}
The evaluation of the distributed RL algorithm was outlined in Section \ref{subsec:gen_episode}. Note that $\epsilon = 0$ and $\mathrm{seq\_len} = 1$ are used during testing, and the hidden and cell state $state\_h_{i,n}$ and $state\_c_{i,n}$ from time slot $n$ are explicitly provided as input to the DQN in the next time slot to emulate a local action-observation history.

Three baselines are used for assessing the performance of the distributed RL algorithm. First, the centralized PF-based BS scheduler presented in Section \ref{sec:PF_scheduler} provides an approximate upper bound on the attainable reward. It is approximate for two reasons: it maximizes $\sum_{n=0}^{L}\gamma^n r[n]$ for $\gamma=1$, and since the centralized scheduler determines the action of all BS's at the beginning of a time slot, it uses the path gains $h_{ij}$ from the previous time slot. However, with $\gamma = 1 - 10^{-6}$ and slow fading coefficient $\alpha = 0.01$, the approximations prove sufficiently accurate. For the reasons outlined in detail in the beginning of Section \ref{sec:dec_ma}, the PF scheduler is not realizable in any practical decentralized deployment in the absence of a central controller.

The second baseline is the ED threshold, which allows a BS to transmit if the received sum of energies of the already transmitting BSs is less than $E_0$ i.e.
BS $i$ transmits if $\sum_{j=1}^{N} \mathcal{E}_{ij}^{\theta_{i}} < E_0$. We employ $E_0 = -72 \ \mathrm{dBm}$ \cite{3gpp.36.889}.

The third baseline, referred to as ``Adaptive ED'' is a configuration adaptive ED threshold. It finds the ED threshold that maximizes $\sum_{n=0}^{L}\gamma^n r[n]$ for the given configuration of UEs from a set of ED thresholds ranging from -32 dBm to -92 dBm. Note that ``Adaptive ED'' is only used to benchmark the performance of the RL algorithm, and is not directly realizable since a BS cannot be cognizant of the configuration of the UEs before transmission. However, one of the strategies considered to make a BS aware of the channel state at the UE is for the UE to respond with a CTS message only if the SINR of the received RTS is greater than a threshold \cite{jamil2015efficient} \cite{lien2016configurable}. In other words, CCA is carried out at the UE also. If the BS does not receive the CTS, it can adapt its ED threshold accordingly. The performance of ``Adaptive ED'' is indicative of such receiver-based LBT mechanisms. This approach however results in high signaling overhead, high latency and wastage of radio resource. The RL algorithm, on the other hand, does not suffer from any of the aforementioned drawbacks. 
\section{Results \& Discussion} \label{sec:Results}
\begin{figure*}
\centering
  \subfloat[Layout 1. $l=100$]{\includegraphics[height=0.25\textheight, width=0.47\textwidth]{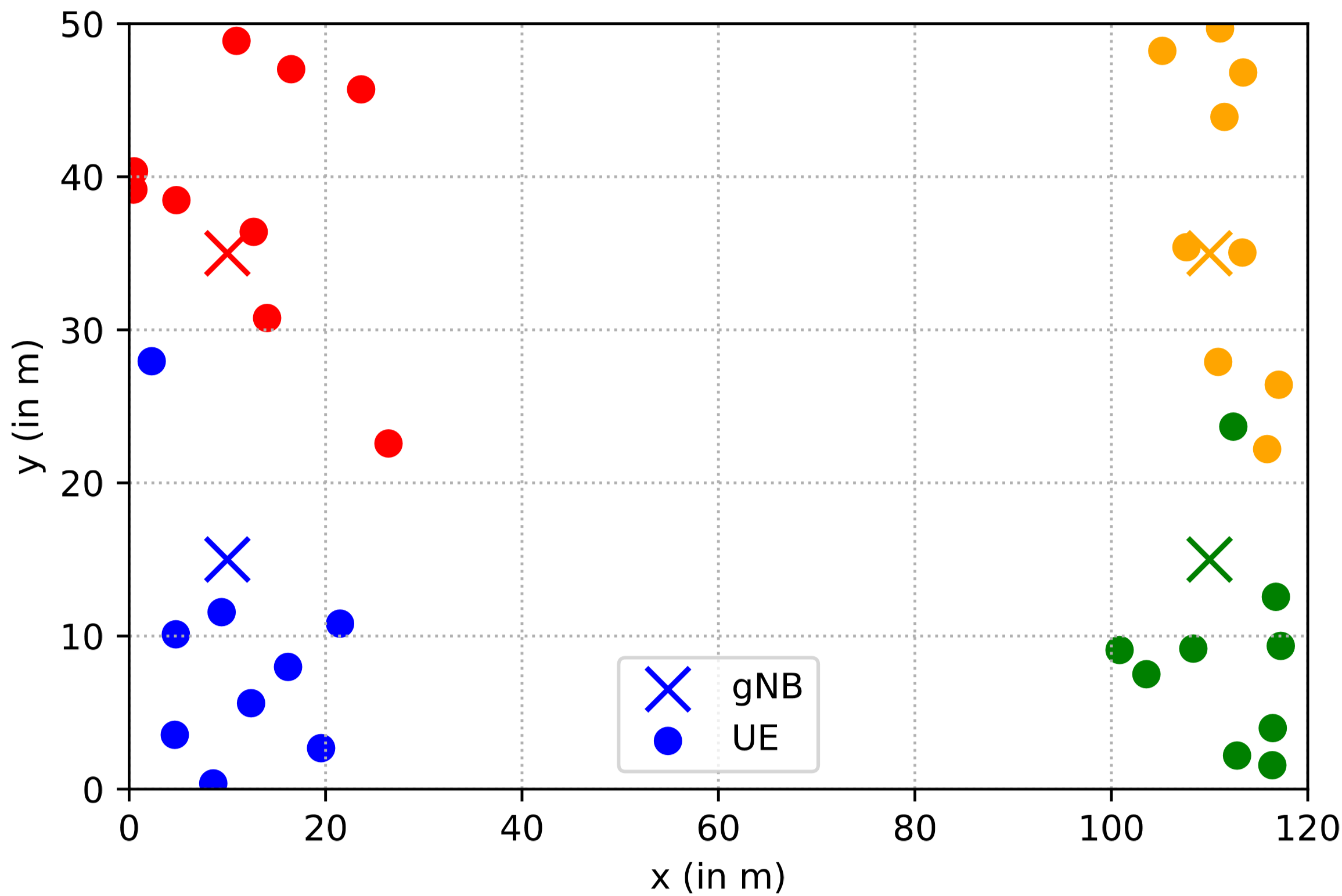}
  \label{fig:layout_1}}
  \hspace{0.2in}
  \subfloat[Layout 2. $l=40$]{\includegraphics[height=0.25\textheight, width=0.47\textwidth]{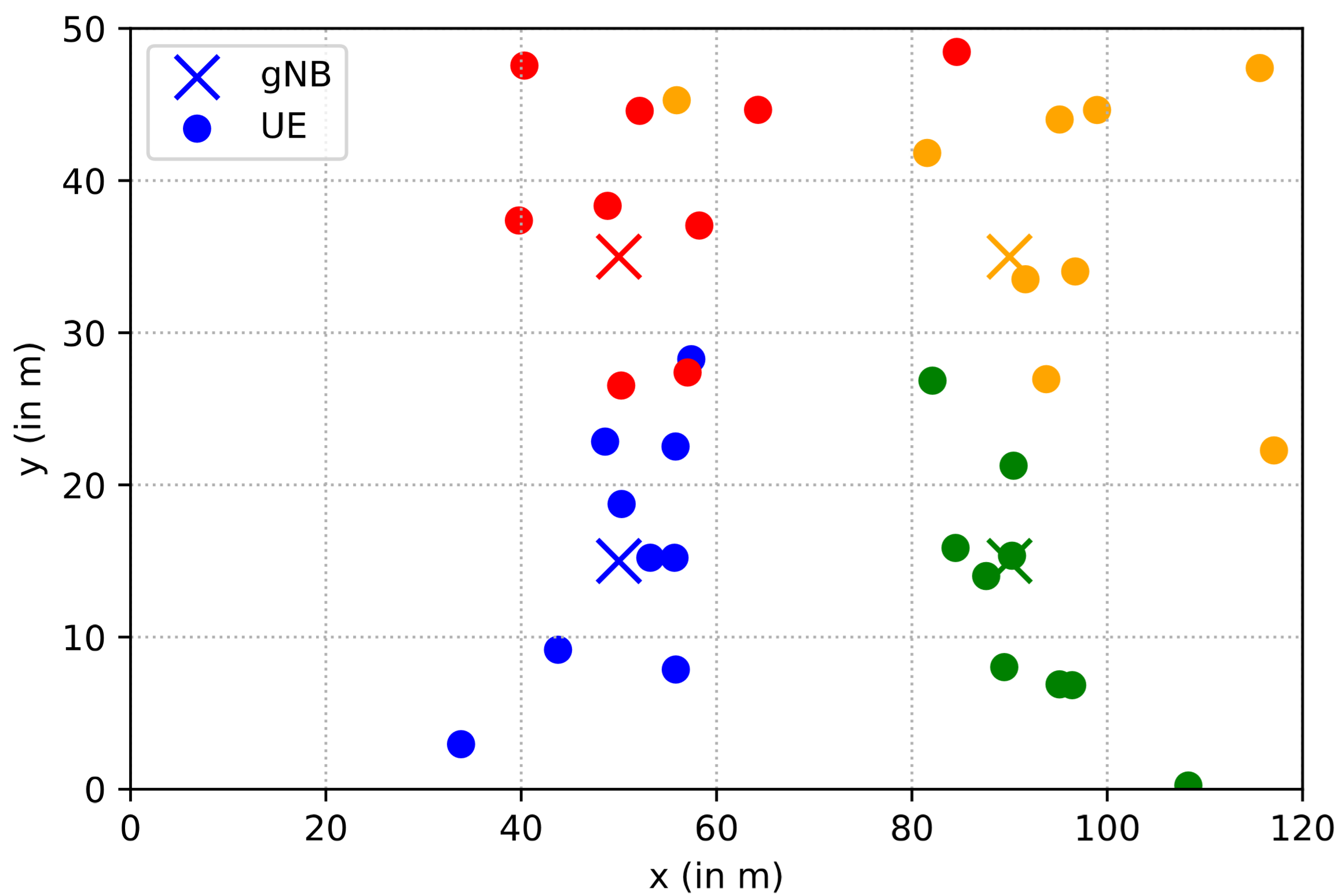}
  \label{fig:layout_2}}
\caption{Two layouts of 4 BS's at the corners of a $l \times 20$ m rectangle. A BS is referred to as gNB in NR terminology.}
\label{fig:layout}
\end{figure*}

We consider 4 BSs lying at corners of a rectangle of breadth 20 m. In \textit{Layout 1} (L1), the length of the rectangle is 100 m, while in \textit{Layout 2} (L2), it is 40 m. Henceforth, we will refer to the position of the BSs as a \textit{Layout}, and the position of the UEs as a configuration. We will be considering only two \textit{Layouts} throughout the simulations, but a large number of configurations. The primary difference between the two layouts is that for most choices of 4 UE's, the inter-BS energies $\mathcal{E}_i$ will accurately reflect the quality of reception in \textit{Layout 1}, but not so in \textit{Layout 2}. This is evident from Fig. \ref{fig:layout_1} where the separation between UE's from different BS's reflects the inter-BS separation more accurately than in Fig. \ref{fig:layout_2}. 

\begin{figure*}
\centering
\subfloat[Layout 1]{\includegraphics[height=0.25\textheight, width=0.47\textwidth]{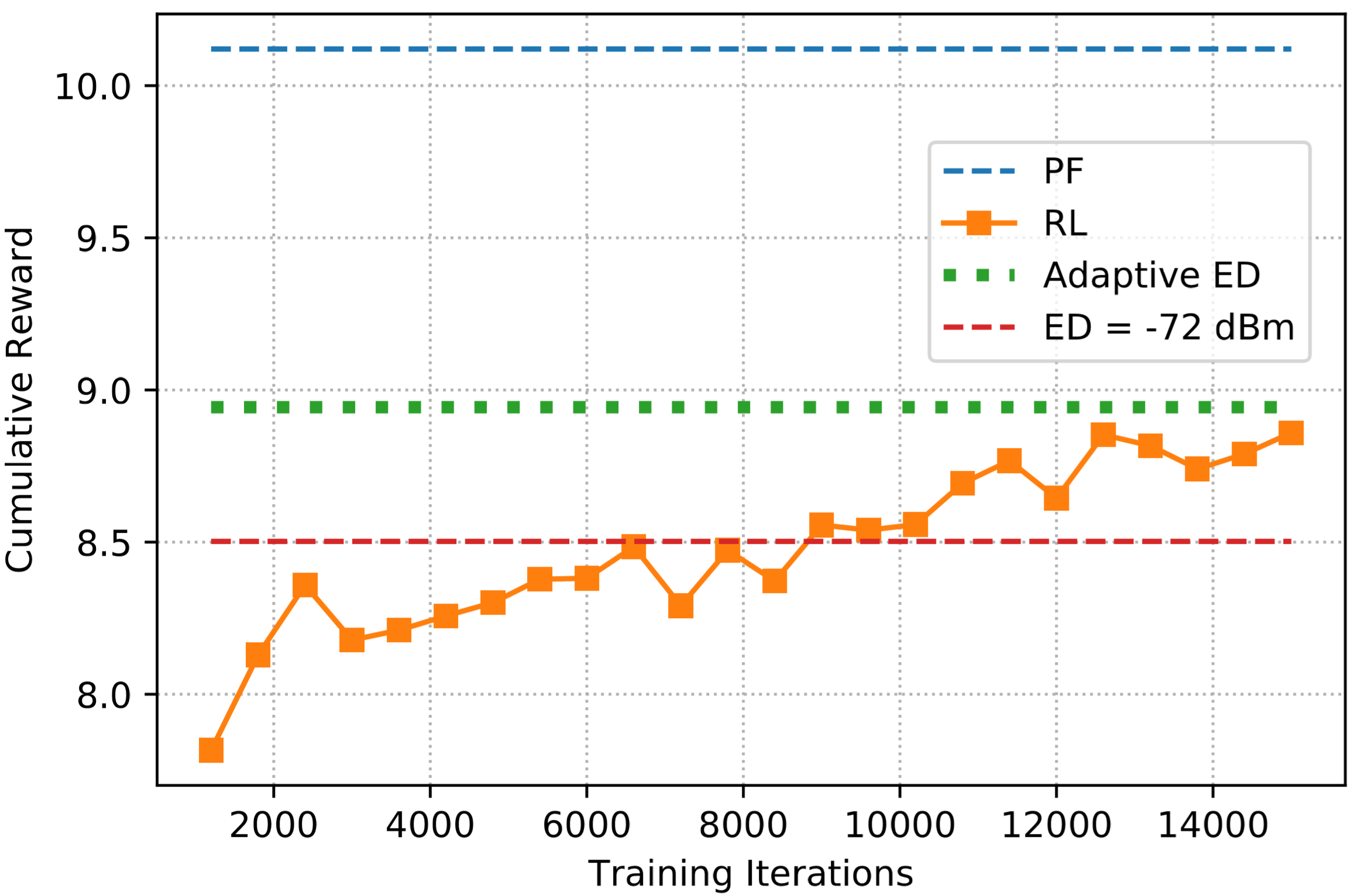}
  \label{fig:reward_iter_layout1}}
  \hspace{0.05in}
  \subfloat[Layout 2]{\includegraphics[height=0.25\textheight, width=0.47\textwidth]{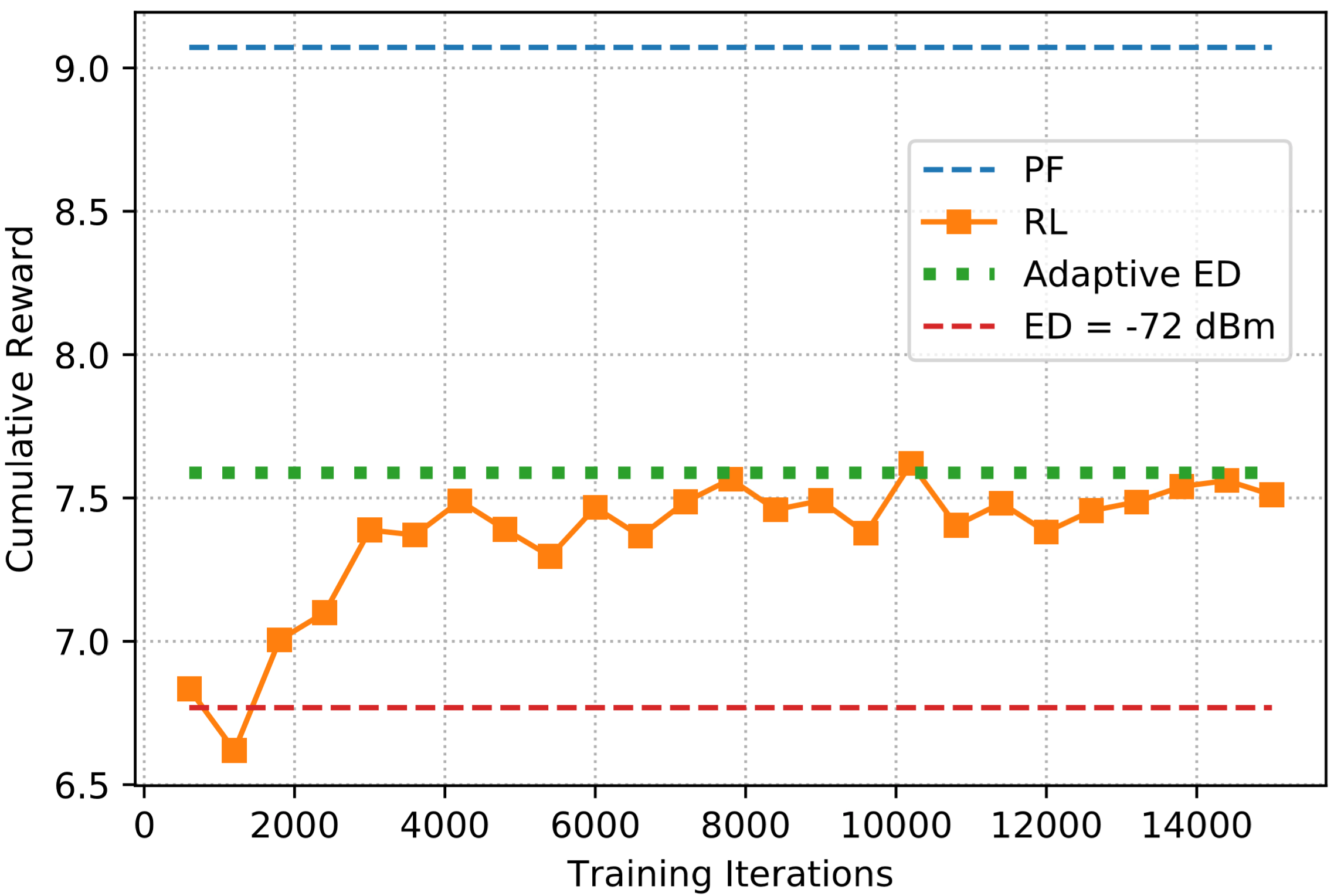}
  \label{fig:reward_iter_layout2}}
\caption{Cumulative Reward evaluated on Validation Set for \textit{Layout 1} and \textit{Layout 2}}
\label{fig:results}
\end{figure*}

The validation curve is shown for Layout 1 and 2 in Fig. \ref{fig:reward_iter_layout1} and \ref{fig:reward_iter_layout2} respectively. It is obtained by evaluating the trained models obtained after every 600 iterations on 10 randomly sampled configurations (not part of the training set) and averaged over 10 realizations of each configuration. The constant benchmarks provided by the PF and ED baselines averaged over the same validation configurations are also plotted. Clearly evident is the increasing gap between the ``Adaptive ED'' and PF baselines as we go from Layout 1 to Layout 2, due to the separation between UE's from different BS's reflecting more precisely the inter-BS separation in Layout 1. As a consequence, Layout 1 vs. Layout 2 also depicts how a single standardised threshold of -72 dBm cannot provide the same degree of fairness in different scenarios. In both cases, the reward accumulated by the RL algorithm gradually merges with the Adaptive ED threshold baseline, and is significantly higher than the average reward obtained using the standardised -72 dBm ED threshold.

For both layouts, the trained models obtained at the end of 15000 training iterations are evaluated on 15 randomly sampled configurations (not part of the training and validation set), with the performance metric averaged over 120 realizations of each configuration (to average over fading and different counter realizations). As in the training phase, unique counters are used for each BS with each counter $\theta_i \in \{0,1,2,3\}$ and $\mathrm{CW} = N = 4$. The same configurations are also used for evaluating the centralized PF-based BS scheduler and ED threshold baselines. The trained RL models, PF and ED algorithms are evaluated on the same fading and counter realizations. The envelope of all the ED threshold curves is used to create the ``Adaptive ED'' threshold curve. The mean reward over all test UE configurations is summarized in Table \ref{tab:mean_reward} for Layout 1 and 2 under the tab UC (Unique Counter).
\begin{table}
    \centering
    \includegraphics[width=2.5in]{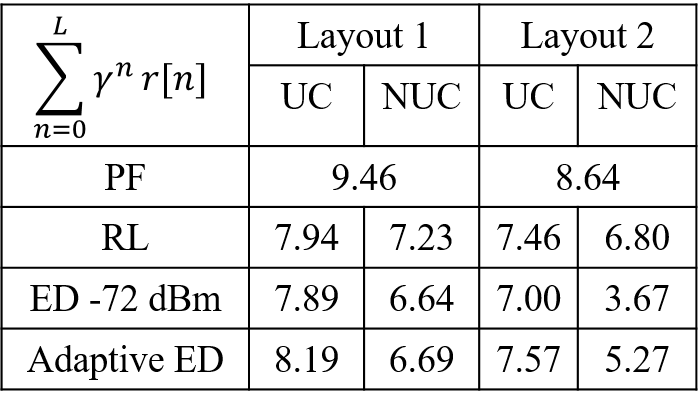}
    \caption{Average Reward for Layout 1 and 2 with CW = 4. UC/NUC is Unique/Non-Unique counter for each BS.}
    \label{tab:mean_reward}
\end{table}

Fig. \ref{fig:results} and Table \ref{tab:mean_reward} highlight the proximity of the RL algorithm to the configuration adaptive ED threshold in terms of maximization of the expected cumulative reward. This is one of the key findings of this paper, since the adaptive ED threshold optimizes the ED threshold for each UE configuration, thus subsuming knowledge of the BS-UE path gains, while the RL algorithm is not provided with this information. On the other hand, the performance of the standardised -72 dBm ED threshold varies significantly depending on the \textit{Layout} as discussed before. We now elaborate on certain key features of the RL algorithm and aspects of medium access protocols that it has the potential to improve.

\subsection{Non-Unique counters} \label{subsec:nuc}
\begin{figure}[!ht]
\centering
    \includegraphics[height=0.25\textheight, width=0.47\textwidth]{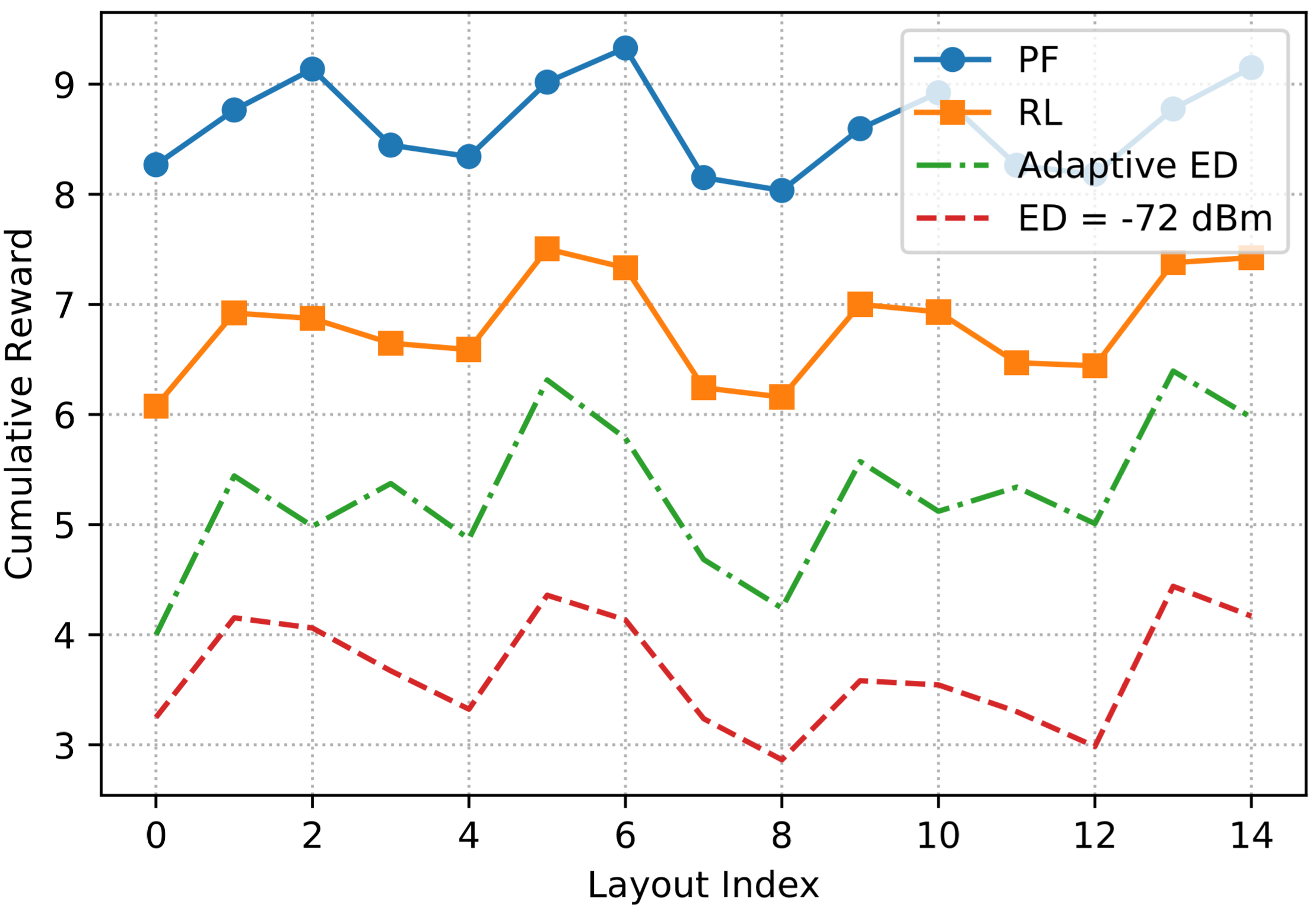}
\caption{Using non-unique counters depicts the increased complexity of learnt RL policy as compared to ED threshold.}
\label{fig:reward_layout2_counter}
\end{figure}

Consider employing non-unique counters while evaluating the trained models, i.e. each BS $i$ can receive a random counter $\theta_i \in \{0,1,2,3\}$, with the possibility that $\theta_i = \theta_j$ for $i \neq j$. However, we maintain $\mathrm{CW} = 4$. The average rewards are given in Table \ref{tab:mean_reward} under the tab NUC (Non-Unique Counter), and the results for Layout 2 are depicted in Fig. \ref{fig:reward_layout2_counter} as a function of the Layout Index ranging from 1 to 15. While the average reward earned by the ED threshold of -72 dBm is significantly reduced (7 to 3.67 and 7.89 to 6.64), the RL algorithm is much more robust to the presence of counter collisions (7.46 to 6.80 and 7.94 to 7.23).

It is evident from Fig. \ref{fig:reward_layout2_counter} that RL achieves a significantly higher reward than even the ``Adaptive ED" baseline. RL provides for the application of a state-based policy at each BS $i$ for determining $a_i$, and in addition to the inter-BS energy vector $\mathcal{E}^{\theta_i}_{i}$, this state also contains the average rate $\overline{X}_i$, signal $S_i$ and interference power $I_i$ of the served UE from the previous time slot. Moreover, the LSTM hidden and cell state $state\_h_{i,n}$ and $state\_c_{i,n}$ provide access to the local action-observation history. However, if two or more BSs have the same counter, each of these BS's misses out on the chance to learn about the decision of another to transmit, thus increasing the extent of partial observability in $\mathcal{E}^{\theta_i}_i$. Our results show that RL learns a complex policy that depends on all the input variables in the CON and EOS state definitions, and is not as impacted by a less informative $\mathcal{E}^{\theta_i}_i$ as the ED threshold, which is entirely dependent on $\mathcal{E}^{\theta_i}_i$ for determining $a_i$. Comparing the UC and NUC tabs in Table \ref{tab:mean_reward}, it is evident that RL is impacted by counter collisions, but not to the extent of the ED threshold.

Conventional WiFi systems reduce the probability of counter collision by using a larger CW size of at least 15 \cite{is2012ieee}. An RL based approach can thus reduce the size of the contention window, paving the way for improved resource utilization in data transmission. Training the RL algorithm with counter collisions and larger CWs could further improve the robustness to non-unique counters, and is left for future work. 

\subsection{Length of Energy Vector $\mathcal{E}_{i}^{\theta_{i}}$} \label{subsec:energy_vector_length}
\begin{figure}[!ht]
    \centering
    \includegraphics[height=0.27\textheight, width=0.475\textwidth]{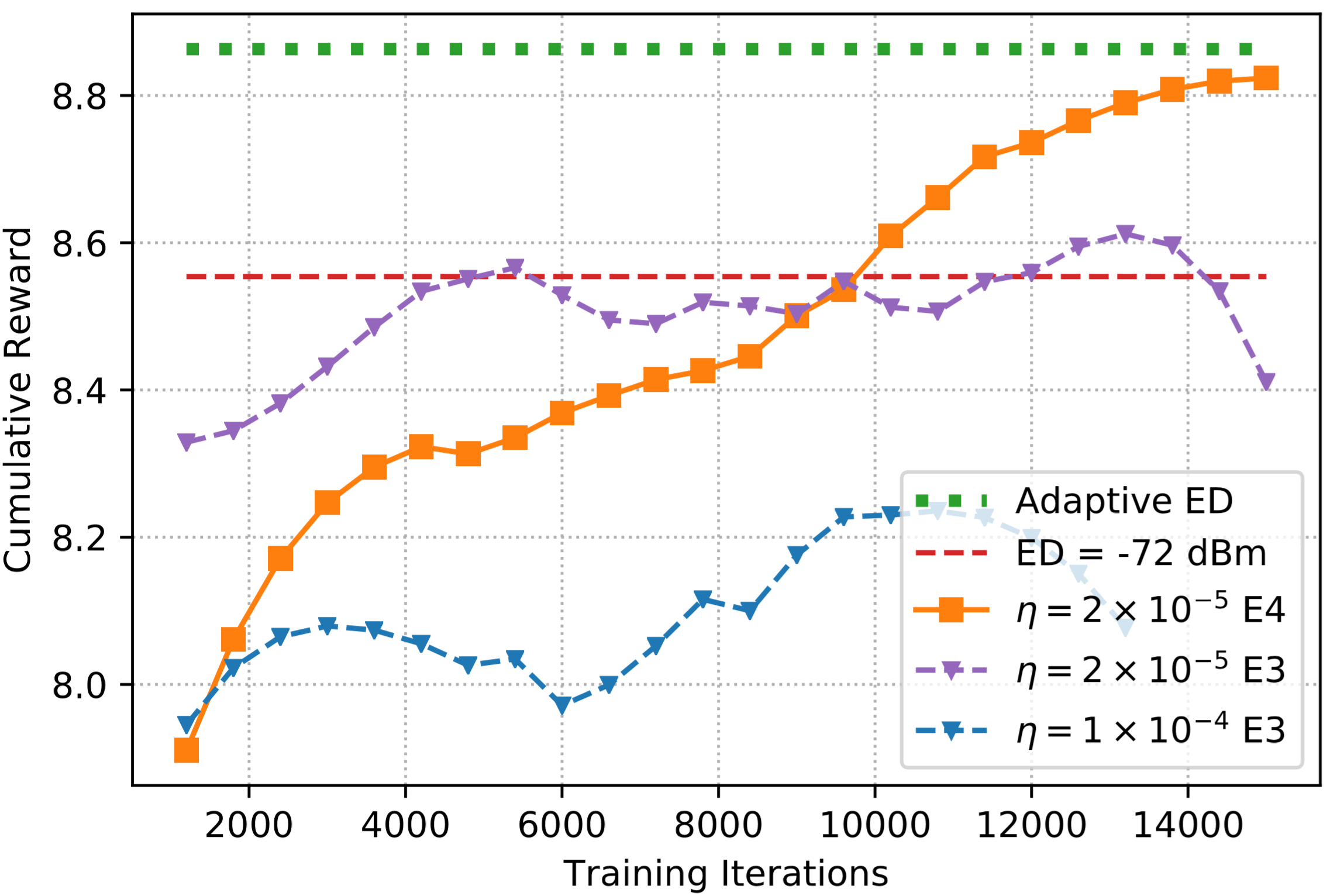}
    \caption{Varying length of $\mathcal{E}_{i}^{\theta_{i}}$ affects training convergence}
    \label{fig:valid_curve_05611_E3_4}
\end{figure}

Since we assume that only DL BS-UE transmissions are sharing the same spectrum, and the rest of the transmissions do not interfere, the term $\mathcal{E}_{ii}^{\theta_{i}}$ would be zero (The BS noise figure used in simulations is small enough to be neglected). A seemingly obvious strategy, that could even bring about some reduction in computational complexity, is to simply delete this entry before inputting $\mathbf{o}^{\mathrm{CON}}_i$ to $\mathcal{Q}^{\mathrm{CON}}_i$. The smoothed validation curves for \textit{Layout 1} with this modification are presented in Fig. \ref{fig:valid_curve_05611_E3_4} with the label E3 (since $\mathcal{E}_{i}^{\theta_{i}}$ was truncated to length $N-1 = 3$), while the original validation curve from Fig. \ref{fig:reward_iter_layout1} is reproduced here with the label E4. For two distinct choices of initial learning rate $\eta$, we are unable to acheive training convergence. A possible explanation could be as follows: Consider again the 3 BS example depicted in Fig. \ref{fig:pic_algo} where BS 1 had the smallest counter and chose to transmit first. If we truncated the length of the energy vector $\mathcal{E}_{i}^{\theta_{i}}$ from $3$ to $2$, then the index of the non-zero term in $\mathcal{E}_{0}^{\theta_{0}}$ and $\mathcal{E}_{2}^{\theta_{2}}$ would be different i.e. the first term would be non-zero in $\mathcal{E}_{0}^{\theta_{0}}$ and the second term in $\mathcal{E}_{2}^{\theta_{2}}$. As a result, in the end-to-end training across agents brought about by the vector $\mathcal{E}_{i}^{\theta_{i}}$ and the common reward $r^{\mathrm{CON}}$, the CON DQN at each BS would be unable to assign an identity to the interfering transmitter. Hence, when evaluating the learnt policy, it is unable to utilize the information contained in $\mathcal{E}_{i}^{\theta_{i}}$ effectively to estimate the interference at the UE in the current time slot.

\subsection{Impact of Fading}
\begin{figure*}
    \centering
    \subfloat[Validation Curve for Layout 2 without fading (in blue)]{\includegraphics[height=0.27\textheight, width=0.475\textwidth]{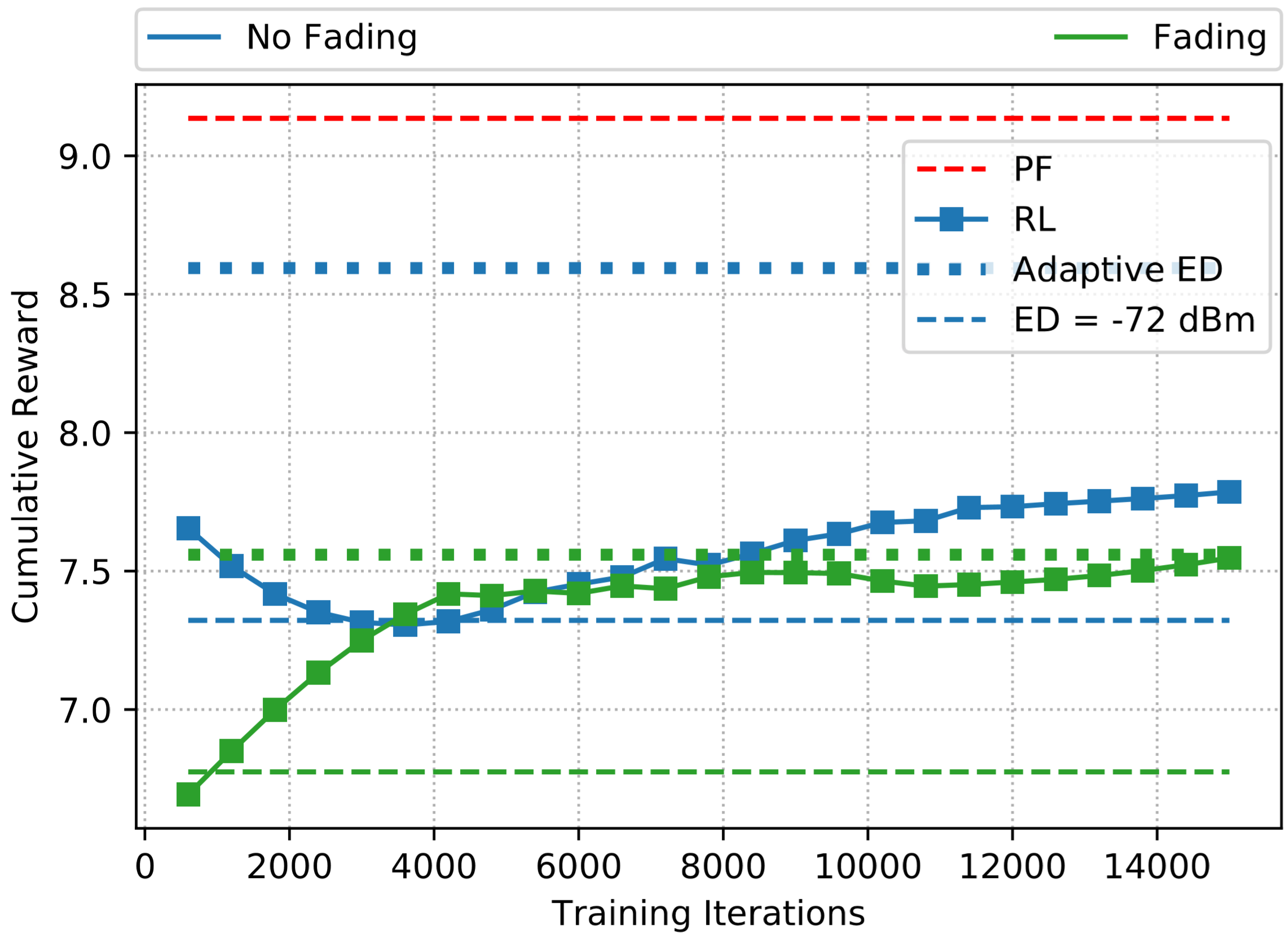}
    \label{fig:valid_curve_f_nf_2479}}
    \hspace{0.1in}
    \subfloat[Validation Curve for Layout 2 with $\alpha = 0.1$.  ]{\includegraphics[height=0.27\textheight, width=0.475\textwidth]{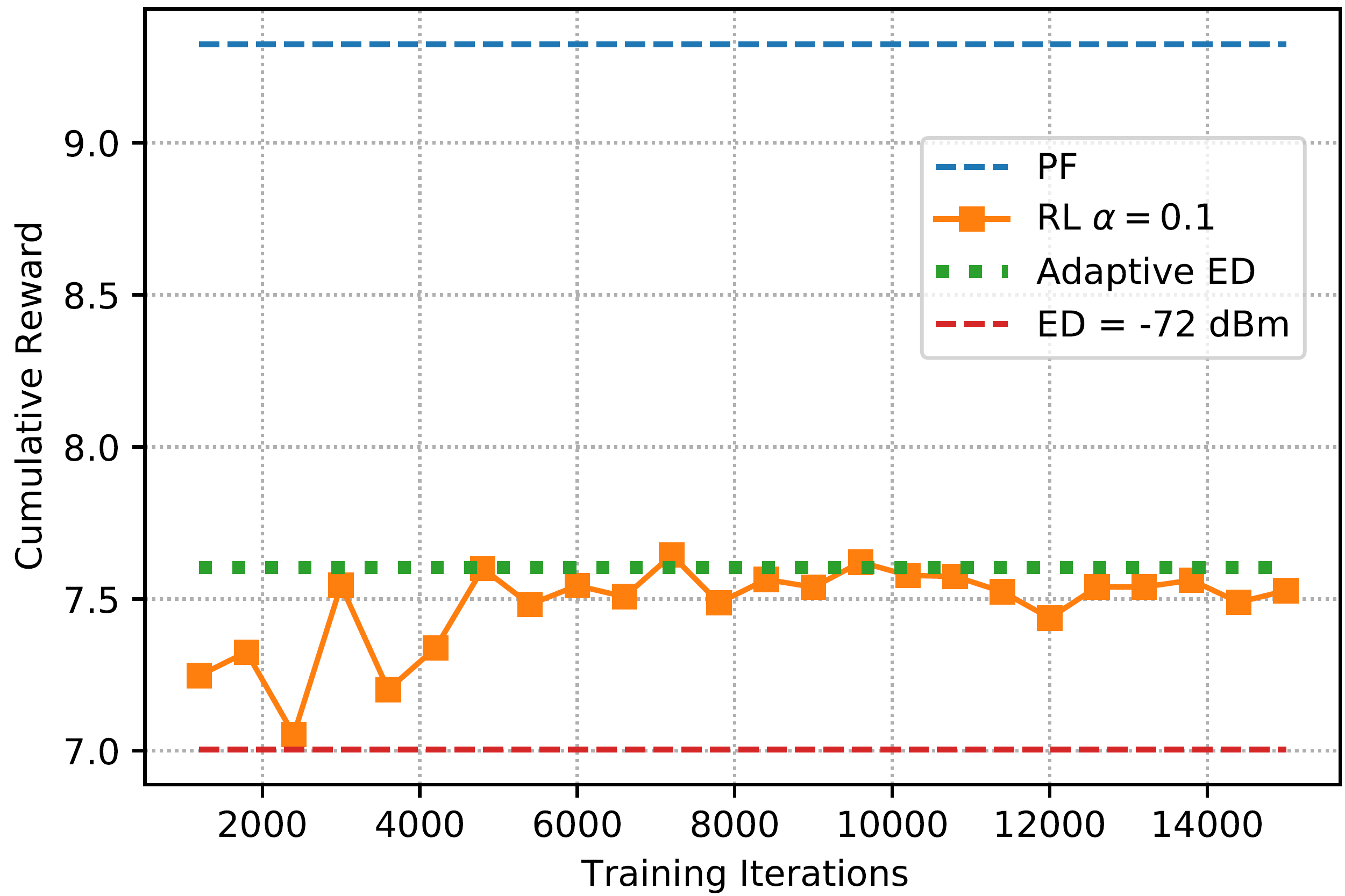}
    \label{fig:valid_curve_alpha_0p1_2479}}
\caption{Our RL-based approach's training convergence and performance is robust to the degree of fading}
\end{figure*}

To investigate the impact fading on all links had on the training convergence of the RL algorithm, we performed two experiments. First, we removed fading both in the training and testing of the algorithm. The smoothed validation curve for \textit{Layout 2} in the absence of fading is shown in Fig. \ref{fig:valid_curve_f_nf_2479} in blue, overlayed with the validation curve in the presence of fading from Fig. \ref{fig:reward_iter_layout2} in green. It appears that the RL algorithm is not significantly impacted by fading, since the $S_i$ and $I_i$ terms in the state definition serve as good indicators of the channel coefficients in a slow fading environment. While both ED baselines in the absence of fading may appear to significantly outperform their fading counterparts, this is simply an artifact of that particular snapshot of the channel gains under ``no fading" being favorable to the ED baselines, and have been shown only for reference. Second, we increased the slow fading coefficient to $\alpha = 0.1$, and the corresponding validation curve for \textit{Layout 2} is depicted in Fig. \ref{fig:valid_curve_alpha_0p1_2479}. Both the PF and ED baselines have also been plotted using $\alpha = 0.1$, and the RL algorithm continues to perform as well as the Adaptive ED baseline. This further validates the observation that the RL algorithm is not noticeably impacted by fading.

\subsection{Online Adaptability} \label{subsec:online}
One of the key advantages of the formulation described thus far is the per-timestep reward structure presented in \eqref{eq:per_ts_rwd}. This reward function is independent of the episode length $L$. Moreover, consider an online deployment where every $f$ time slots, a BS receives messaging on backhaul links from other BS that could contain the average rate of the remaining UEs and channel state information (CSI) that could aid in determining $\{g_{ij}\}$ contributing to interference at the UE served. Then the BS could utilize this information to update the weights of its DQN's by simply using the same reward function from \eqref{eq:per_ts_rwd} to calculate the labels $L^{\mathrm{EOS}}_i$ and $L^{\mathrm{CON}}_i$ as given by \eqref{eq:l_eos} and \eqref{eq:l_con}. This follows from the observation that in the absence of an update, if we consider $R_j[n]$ to equal its expected value $\overline{X}_j[n-1]$, then
\begin{equation}
    \log\Bigg((1-1/B)\Big(1 + \frac{\overline{X}_j[n-1]}{(B-1)\overline{X}_j[n-1]}\Big)\Bigg) = 0.
\end{equation}
Hence the cumulative reward $\approx \sum_n r[n]$ for $\gamma \rightarrow 1$ remains unaffected by missed updates, and we continue to optimize for long term PF.

\section{Conclusions and Future Directions} \label{sec:conc}
Transmitter side MAC mechanisms such as the ED threshold algorithm, adopted as the the starting point of the design of LAA \cite{3gpp.36.889} and the unlicensed 6 GHz band in NR-U\cite{3gpp.38.889}, incorrectly assume that interference sensed at the BS is representative of the quality of reception at the UE. To develop a state-based MAC policy, we first formulated Medium Access as a \textit{DEC-POMDP} and incorporated contention to derive a 2-state transition diagram for the BS in each time slot. We then presented a distributed deep $\mathcal{Q}$-learning algorithm that utilized an LSTM layer in each DQN to capture the local action-observation history in combination with message passing to provide for end-to-end training across BSs. The algorithm developed in this paper jointly utilizes the information from LBT-based spectrum sensing at the BS along with the average rate, signal and interference power seen by the UE it serves to determine whether the BS will transmit in the designated time slot. Moreover, the DQN’s at each BS can be trained offline and refined periodically online owing to the structure of the per-reward timestep, providing for online adaptability. These features make our approach quite suitable for deployment in actual BSs.

Utilizing a centralized training procedure with decentralized execution, the distributed RL algorithm was found to match the performance of a configuration adaptive ED threshold. Moreover, we showed that it has the potential to reduce the size of the contention window, providing for improved resource utilization in data transmission, and its training convergence is minimally impacted by the presence of fading on all links.

One of the drawbacks of the approach we have presented is the need for deploying a different DQN at each BS. The current framework has no parameter sharing among DQNs at different BSs. Ideally, we would want to train a single DQN that can be deployed at any BS, and refine it online via periodic updates. Moreover, there is still significant room for improvement in terms of the gap from the centralized PF-based BS scheduler, and policy gradient methods \cite{sutton2018reinforcement} could help. With a view to the design of a learning based BS, the framework developed in this paper has the potential to be applied in a variety of decision-making problems, including rate control, beam selection for scheduling, coordinated scheduling and channel selection in the frequency domain. In essence, these extensions tremendously increase the dimensionality of the output action space, from a simple transmit Yes/No decision to a choice of MCS, beamformer, user and subcarrier. With machine learning (ML) envisioned to be a key enabler for 6G \cite{dang2020should}, distributed learning designs of spectrum sharing mechanisms for the increasing amounts of unlicensed spectrum being made available should prove to be of paramount importance.

\appendix
\subsection{Proof of Eqn \eqref{eq:per_ts_rwd}} \label{subsec:proof_per_ts_rwd}
\noindent In order to satisfy  \eqref{eq:rwd_approx}, let us rewrite \eqref{eq:avg_rate} as follows, with $n' = n + 1$
\begin{equation} 
\label{eq:log_avg_rate}
    \overline{X}_j[n'] = \Big(1-\frac{1}{B}\Big)\overline{X}_j[n]\Big(1 + \frac{R_j[n]}{(B-1)\overline{X}_j[n]}\Big).
\end{equation}
Observe that by taking $\log$ on both sides, using \eqref{eq:avg_rate} to rewrite the outer $\overline{X}_j[n]$ in terms of $\overline{X}_j[n-1]$ and so on until we reach $\overline{X}_j[0]$ in \eqref{eq:log_avg_rate}, and using the $\log$ function to split the product as a sum, we can obtain \begin{equation}
   \log(\overline{X}_j[L]) = \log({X}_j[0]) + \sum_{n=1}^{L} r_j[n],
\end{equation}
where $r_j[n]$ denotes the contribution of each UE to the reward $r[n]$ and is given by \eqref{eq:per_ts_rwd}. Hence, we have \begin{equation}
    U(\mathbf{\overline{X}}[L]) = \sum_{j=1}^{N} \log({X}_j[0]) + \sum_{j=1}^{N} \sum_{n=1}^{L} r_j[n].
\end{equation}

\bibliographystyle{IEEEtran}
\bibliography{bibtex.bib}

\end{document}